\documentclass[final]{arxiv}
\pdfoutput=1

\usepackage{natbib}
\usepackage{url}
\RequirePackage[OT1]{fontenc}
\RequirePackage{graphicx}
\RequirePackage{amsthm}
\RequirePackage{amssymb}
\RequirePackage[cmex10]{amsmath}
\usepackage[colorlinks,citecolor=blue,urlcolor=blue]{hyperref}

\usepackage{comment}

\usepackage{mathpazo}
\usepackage[notcite,notref]{showkeys} 
\usepackage{todonotes}
\usepackage[normalem]{ulem}

\startlocaldefs
\theoremstyle{plain}
\newtheorem{theorem}{Theorem}

\newtheorem{Lemma}[theorem]{Lemma}

\newtheorem{Definition}[theorem]{Definition}

\DeclareMathAlphabet{\mathbbold}{U}{bbold}{m}{n}
\def\0{{\mathbbold0}}
\def\1{{\mathbbold1}} 
\newcommand{\commentout}[1]{} 
\renewcommand{\arraystretch}{1.3}
\def\reals{{\mathbb R}}

\def\T{^{\top}} 
\def\N{{\mathcal N}} 

\def\H{{H}} 
\def\la{{\underline{\alpha}}} 
\def\ua{{\overline{\alpha}}} 
\def\HP{{\overline{P}}} 
\def\HQ{{\overline{Q}}} 
\def\lx{{\underline{x}}} 
\def\ux{{\overline{x}}} 
\def\ly{{\underline{y}}} 
\def\uy{{\overline{y}}} 
\def\llam{{\underline{\lambda}}} 
\def\ulam{{\overline{\lambda}}} 
\newdimen\einr
\def\myitem#1{\par\hangafter=1\hangindent=\einr\noindent
  \hbox to\einr{\ignorespaces#1\hfill}\ignorespaces}
\def\tombstone{\hbox{\lower.4pt\vbox{\hrule\hbox{\vrule
  \kern7.6pt\vrule height7.6pt}\hrule}\kern.5pt}}
\def\myendproof{\hfill\strut\nobreak\hfill\tombstone\par\medbreak}
\newcommand{\mini}{\mathop{\hbox{\rm minimize }}}
\newcommand{\maxi}{\mathop{\hbox{\rm maximize }}}
\newcommand{\subj}{\hbox{\rm subject to }}
\def\Primal{P} 
\def\OPT{\textit{OptFace}}
\def\SLmin{\textit{SL}^{\min}}
\def\SLmax{\textit{SL}^{\max}}
\def\BRmin{\textit{BR}^{\min}}
\def\BRmax{\textit{BR}^{\max}}
\def\Qmin{\textit{\Primal}^{\,\min}}
\def\Qmax{\textit{\Primal}^{\,\max}}
\def\zerom{{\bf0}}
\def\zeron{{\0}}
\def\onem{{\bf1}}
\def\onen{\1} 

\newcounter{alg}

\def\TAB{\leavevmode\hbox to 2em{\hfill}\ignorespaces}
\def\NL{\par\leavevmode\refstepcounter{alg}\noindent
   \hbox to 1.5em{\hfill\footnotesize\arabic{alg}}\hbox
   to 2em{\hfill}\ignorespaces}

\def\PROCNAMEFONT{\textsc}

\def\STAR#1{{#1^*}}
\def\BINSEARCH{\PROCNAMEFONT{BinSearch}} 

\def\myproof{\proof{Proof.}}
\def\myendproof{\hfill\strut\nobreak\hfill\Halmos\par\endproof} 

\def\myproof{\proof}
\endlocaldefs
\def\Halmos{\mbox{\quad$\square$}}

\begin{document}

\begin{frontmatter}
\title{Fast Algorithms for Rank-1 Bimatrix Games}
\runtitle{Fast Algorithms for Rank-1 Bimatrix Games}

\begin{aug}
\author{\fnms{Bharat}
\snm{Adsul}\thanksref{a1}\ead
[label=e1]{adsul@cse.iitb.ac.in}\ead
[label=u1,url]{www.cse.iitb.ac.in/page14}} 
\author{\fnms{Jugal}
\snm{Garg}\thanksref{a2}\ead[label=e2]{jugal@illinois.edu}\ead
[label=u2,url]{www.jugal.ise.illinois.edu}} 
\author{\fnms{Ruta} \snm{Mehta}\thanksref{a2}\ead[label=e3]{rutamehta@illinois.edu}\ead[label=u3,url]{www.rutamehta.cs.illinois.edu}}
\author{\fnms{Milind}
\snm{Sohoni}\thanksref{a1}\ead[label=e4]{sohoni@cse.iitb.ac.in}\ead
[label=u4,url]{www.cse.iitb.ac.in/page14}} 
\author{\fnms{Bernhard} \snm{von Stengel}\thanksref{a3}
\ead[label=e5]{b.von-stengel@lse.ac.uk}
\ead[label=u5,url]{www.maths.lse.ac.uk/Personal/stengel/}}

\runauthor{Adsul, Garg, Mehta, Sohoni and von Stengel}

\address[a1]{Department of Computer Science and Engineering,
Indian Institute of Technology Bombay, Powai, Mumbai 400
076, India,
\printead{e1,u1}\\
\printead{e4,u4}
}
\address[a2]{University of Illinois at Urbana-Champaign, Urbana, IL 61801, USA, 
\\
\printead{e2,u2}\\
\printead{e3,u3}
}

\address[a3]{Department of Mathematics, London School of Economics, London WC2A 2AE, United Kingdom,
\printead{e5};
\printead{u5}
}
\end{aug}

\begin{abstract}
The rank of a bimatrix game is the matrix rank of the sum of
the two payoff matrices.
This paper comprehensively analyzes games of rank one, and
shows the following:
(1)
For a game of rank~$r$, the set of its Nash equilibria is the
intersection of a generically one-dimensional set of
equilibria of parameterized games of rank $r-1$ with a
hyperplane.
(2)
One equilibrium of a rank-1 game can be found in polynomial
time.
(3)
All equilibria of a rank-1 game can be found by following
a piecewise linear path.
In contrast, such a path-following method finds only one
equilibrium of a bimatrix game.
(4)
The number of equilibria of a rank-1 game may be
exponential.
(5)
There is a homeomorphism between the space of bimatrix games
and their equilibrium correspondence that preserves rank.
It is a variation of the homeomorphism used for the concept
of strategic stability of an equilibrium component.

\end{abstract}


\end{frontmatter}

\section{Introduction}
\label{s-intro}

Non-cooperative games are basic economic models.
The main concept to analyze them is Nash equilibrium, which
recommends to each player a (typically randomized) strategy
that is optimal for that player if the other players follow
their recommendations.
In order to give such a recommendation, a Nash equilibrium
must be found by some method (including any adjustment
process).
For larger games this requires computer algorithms.
We consider bimatrix games, which are two-player games in
strategic form.
The algorithm by \citet{LH} finds one equilibrium of a
bimatrix game.
Finding all equilibria is feasible only for small games
because of the exponential number of mixed strategies that
typically need to be checked for the equilibrium property
\citep{ARSvS}.

\citet{KT2010} introduced a hierarchy of bimatrix games
based on the matrix rank of the sum of the two payoff
matrices.
Games of rank~0 are zero-sum games, which can be solved by
linear programming.
This paper comprehensively studies games of rank~1.
Rank-1 games are economically more interesting than zero-sum
games, by allowing a ``multiplicative'' interaction in
addition to an arbitrary zero-sum component (discussed
further in Section~\ref{s-conclusions}). 
We will show that, like general bimatrix games, they can
have exponentially many disjoint equilibria.
On the other hand, as our main results show, they are
\textit{computationally tractable}:
One equilibrium of a {rank-1} game can be found
fast (in polynomial time), and finding all equilibria
takes comparable time to finding a single equilibrium of a
general bimatrix game.
Large rank-1 games are therefore attractive as detailed
models of interaction, on a similar scale to, but more
general than, zero-sum games.
Rank-1 bimatrix games and their computational analysis
should therefore become a new tool in economic modeling.

The computational complexity (required running time) of
computing a Nash equilibrium of a game has received
substantial interest in the last two decades.
A computational problem is considered tractable if it can be
solved in polynomial time.
\citet{SvS2006} showed that the algorithm by \citet{LH}
may have exponential running time.
(Their examples require carefully constructed matrices,
comparable to linear programs where the simplex algorithm,
which otherwise works well in practice, has exponential
running time, see \citealp{KleeMinty}.)
The path-following Lemke--Howson algorithm implies that
finding an equilibrium of a bimatrix game belongs to
the complexity class PPAD defined by \citet{Papa1994}.
PPAD describes certain computational problems where the
existence of a solution is known, and the problem is to find
one explicit solution.
(In contrast, the better known complexity class NP applies to
decision problems, which are problems that have a ``yes'' or
``no'' answer.)
Other problems in PPAD include the computation of an
approximate Brouwer fixed point, related
problems in economics such as market
equilibria \citep{VY2001}, and the computation of an
approximate Nash equilibrium of a game with many
players.
(In games with three or more players, unlike in two-player
games, the mixed strategy probabilities in a Nash
equilibrium may be irrational numbers.
A suitable concept for such games is approximate Nash
equilibrium, and finding an exact Nash equilibrium is an
even harder computational problem, see \citealp{EY10}.)
A celebrated result is that all problems in PPAD can be
reduced to finding a Nash equilibrium in a bimatrix game,
which makes this problem ``PPAD-complete''
\citep{ChenDeng2006,CDT2009,DGP2009}.
No polynomial-time algorithm for finding a Nash equilibrium
of a general bimatrix game is known.

\citet{KT2010} describe an algorithm to find
$\varepsilon$-approximate Nash equilibria in games of fixed
rank, with running time that is polynomial in
$1/\varepsilon$ and the input length, but exponential in the
rank. 
In the present paper, we prove that an \textit{exact} Nash
equilibrium of a rank-1 game can be found in polynomial
time.
However, we also show that a rank-1 game may have
exponentially many equilibria.
Moreover, games of higher fixed rank~$r$ are PPAD-hard and
thus as computationally difficult as general bimatrix games;
this has been shown by \citet{Mehta} for $r\ge3$ and 
is claimed to hold for $r=2$ \citep{COPY}.
In the context of the ``rank'' hierarchy, rank-1 games are
therefore the most complex type of games that are expected
to be computationally tractable.

Section~\ref{s-prelims} states the notation and preliminary
results used in this paper, and compares our approach with
the work of \citet{T2009}.
In Section~\ref{s-rankreduce}, we show that the set of
equilibria of a game of rank~$r$ is the intersection of a
hyperplane with a set of equilibria of parameterized games
of rank $r-1$.
When $r=1$, these are parameterized zero-sum games whose
equilibria are the solutions to a parameterized linear
program (LP).
In order to deal with possibly degenerate games which are
awkward to handle with pivoting methods, we recall relevant
results from \citet{AdlerM1992} in Section~\ref{s-paraLP}.
The intersection with the hyperplane gives rise to a
polynomial-time binary search for one equilibrium of a
rank-1 game, explained in Section~\ref{s-binsearch}.
In Section~\ref{s-enum}, we describe completely the set of
all Nash equilibria of a rank-1 game, and outline a
corresponding equilibrium enumeration method.

Section~\ref{s-example} describes an example
(which may be useful to consult in between) 
that illustrates our main results, and a second example that
shows that binary search fails in general for games of
rank~2 or higher.
A construction of rank-1 games with exponentially many
equilibria is shown in Section~\ref{s-expo}.
In Section~\ref{s-structure}, we describe a variant of the
structure theorem of \citet{KM}, which is important for the
concept of strategic stability of an equilibrium component.
We introduce a new homeomorphism between the space of
bimatrix games and its equilibrium correspondence.
This homeomorphism preserves the sum of the payoff matrices,
and hence the rank of the games.
In the concluding Section~\ref{s-conclusions}, we present a
tentative example of an economic model based on rank-1
games, and note some open questions.

A preliminary version of our work was published in STOC 2011
\citep{AGMS2011}, and the result of Section~\ref{s-expo} in
\cite{vS2012}.
The mathematical development in the present paper is almost
entirely new in all parts.

\section{Bimatrix games and best responses}
\label{s-prelims}

In this section we state our notation for bimatrix games and
recall the ``complementarity'' characterization of Nash
equilibria in terms of suitable polyhedra.
We also briefly compare our approach with~\citet{T2009}.

We use the following notation.
The transpose of a matrix $C$ is written $C\T$.
All vectors are column vectors, so if $x\in \reals^m$
then $x$ is an $m\times 1$ matrix and $x\T$ is the
corresponding row vector in $\reals^{1\times m}$.
In matrix products, scalars are treated like $1\times 1$
matrices.
Let $\0$ and $\1$ be vectors with all components equal to 0
and~1, respectively, their dimension depending on the context.
Inequalities like $x\ge\0$ hold for all components.
The components of a vector $x\in \reals^m$ are
$x_1,\ldots,x_m$.

For $c\in\reals^k$ and $\gamma\in\reals$, a
\textit{hyperplane} is
of the form $\{z\in\reals^k\mid {c\T z=\gamma}\}$,
and a \textit{halfspace}
of the form $\{z\in\reals^k\mid c\T z\le\gamma\}$.
A \textit{polyhedron} is an intersection of finitely many
halfspaces, and called a \textit{polytope} if it is
bounded.
A \textit{face} of a polyhedron $P$ is of the form
$P\cap \{z\in\reals^k\mid c\T z=\gamma\}$ where
$P\subseteq\{z\in\reals^k\mid c\T z\le\gamma\}$.
It can be shown that any face of $P$ can be obtained by
turning some of the inequalities that define $P$ into
equalities \cite[Section~8.3]{Schrijver}.
If a face of $P$ consists of a single point, it is called
a \textit{vertex} of $P$.
If $S\subseteq X\times Y$ for sets $S,X,Y$, then
$\{ x\in X \mid (x,y)\in S$ for some $y\in Y\,\}$
is called the \textit{projection} of $S$ on~$X$,
also written as $\{ x \mid (x,y)\in S\,\}$.

A \textit{bimatrix game} is a pair $(A,B)$ of $m\times n$
matrices with rows as pure strategies of player~1 and columns
as pure strategies of player~2.
The players simultaneously choose their pure strategies,
with the corresponding
entry of $A$ as payoff to player~1 and of $B$ to player~2.
The sets $X$ and $Y$ of \textit{mixed} (that is, randomized)
strategies of player~1 and player~2 are given by
\begin{equation}
\label{XY}
X=\{x\in\reals^m\mid x\ge\0,~\1\T x=1\},
\quad
Y=\{y\in\reals^n\mid y\ge\0,~\1\T y=1\}.
\end{equation}
For the mixed strategy pair $(x,y)\in X\times Y$,
the expected payoffs to the two players 
are $x\T A y$ and $x\T B y$, respectively.
A \textit{best response} $x$ of player~1 against $y$ maximizes
his expected payoff $x\T Ay$, and a best response $y$ of
player~2 against $x$ maximizes her expected payoff $x\T By$.
A \textit{Nash equilibrium}~(NE) is a pair of mutual best
responses.

Consider mixed strategies $x\in X$ and $y\in Y$.
If $x$ is a best response to~$y$, then its expected payoff
$x\T Ay$ is clearly at least the payoff $(Ay)_i$ for any
pure strategy~$i$ of player~1.
Moreover, $x$ is a best response to $y$ if and only if any
pure strategy $i$ in the \textit{support} of~$x$ (that is,
where $x_i>0$) is a pure best response to~$y$
\citep{Nash1951}.
The following lemma, due to~\cite{mangasarian1964}, states
this ``best-response condition'' in terms of suitable
polyhedra.

\begin{Lemma}
\label{l-nechar}
Let $(A,B)$ be an $m\times n$ bimatrix game.
Consider the polyhedra 
\begin{equation}
\label{HPQ}
\begin{array}{rclcrcl}
\HP&=&\{(x,v)\in X\times \reals& \mid& B\T x&\le&\1v\,\},
\\ 
\HQ&=&\{(y,u)\in Y\times \reals& \mid& Ay&\le&\1u\,\}.
\end{array}
\end{equation}
Let $(x,y) \in X\times Y$.
Then $x$ is a best response to $y$ if and only if
$(y,u)\in\HQ$ 
and for all rows $i$ 
\begin{equation}
\label{xbr}
x_i=0
\quad\hbox{or}\quad
(Ay)_i=u
\qquad (1\le i\le m),
\end{equation}
and $y$ is a best response to $x$ if and only if
$(x,v)\in\HP$
and for all columns $j$
\begin{equation}
\label{ybr}
y_j=0
\quad\hbox{or}\quad
(B\T x)_j=v
\qquad (1\le j\le n).
\end{equation}
If both conditions hold, then $u$ and $v$ are the unique
payoffs to player 1 and 2 in the Nash equilibrium $(x,y)$. 
\end{Lemma}

A bimatrix game is \textit{degenerate} if there is a mixed
strategy that has more pure best responses than the size of
its support \citep{vS2002}.
A degenerate game may have infinite sets of equilibria.
They can be described by suitable faces of of $\HP$ and
$\HQ$, as explained further in Section~\ref{s-enum}.
Our analysis applies to general games that may be degenerate.

The object of study of our paper are bimatrix games of fixed
\textit{rank}, introduced by \citet{KT2010}.
They generalize zero-sum games, which are games of rank zero.

\begin{Definition}
\label{d-rank}
The \textit{rank} of a bimatrix game $(A,B)$ is the matrix
rank of ${A+B}$.
\end{Definition}

For comparison of our approach with \cite{T2009}, we
consider a quadratic program, due to \cite{MangasarianStone1964},
that captures the NE of $(A,B)$.

\begin{Lemma}
\label{l-QP}
The strategy pair $(x,y)$ is a Nash equilibrium of $(A,B)$
if and only if it is a solution to 
\begin{equation}
\label{QP}
\maxi_{x,y,u,v} x\T(A+B)y - u -v 
\quad \subj\quad
(x,v) \in \HP, \quad  (y,u) \in \HQ.
\end{equation}
The optimum value of $(\ref{QP})$ is zero, with $u=x\T Ay$
and $v=x\T By$.
\end{Lemma}

\myproof
Consider any solution to (\ref{QP}).
Then $v$ is at least the best-response
payoff of player~2 against~$x$ because $(x,v)\in \HP$,
and $u$ is at least the best-response payoff of player~1
against~$y$ because $(y,u)\in \HQ$.
Hence, $x\T(A+B)y -u - v\le 0$.
Furthermore, (\ref{xbr}) and (\ref{ybr}) imply that
$x\T(A+B)y -u -v$ is zero if and only if $(x,y)$ is a NE, in
which case $u=x\T Ay$ and $v=x\T By$.
\myendproof

The quadratic program \eqref{QP} shows the importance of the
rank of the matrix $A+B$.
For zero-sum games, the rank of $A+B$ is zero and (\ref{QP})
is a linear program, a well-known fact \citep{Dantzig1963}.
For a rank-1 game $(A,B)$ with $A+B=ab\T$, the bilinear term
$x\T(A+B)y$ in the objective function becomes the product 
$(x\T a)(b\T y)$ of two linear terms.
The resulting optimization problem is called a
\textit{linear multiplicative program}.
Solving a general linear multiplicative program is NP-hard
\citep{M96}.

Consider a rank-1 game $(A,B)$ where $A+B=ab\T$.
Similar to parametric simplex methods for solving
linear multiplicative programs \citep{konno1991parametric},
\citet{T2009} describes an algorithm to enumerate all
equilibria of $(A,B)$.
For a real parameter~$\xi$, he considers the parameterized LP
\begin{equation}
\label{theo}
\maxi_{x,y,u,v} x\T a\xi - u -v 
\quad \subj\quad
(x,v) \in \HP, \quad  (y,u) \in \HQ, 
\quad b\T y =\xi\,.
\end{equation}
In any solution to (\ref{theo}), $x\T a\xi = x\T a b\T y =
x\T(A+B)y$.
Hence, by Lemma~\ref{l-QP}, any optimal solution to
(\ref{theo}) is an equilibrium of $(A,B)$ if and only its
optimum is zero. 
Moreover, $b\T y=\xi$ implies that $\xi$ is a convex
combination of the components $b_1,\ldots,b_n$ of $b$,
so that one can restrict $\xi$ to the interval 
$[\min\{b_1,\ldots,b_n\},$ $\max\{b_1,\ldots,b_n\}]$.
By partitioning this interval into segments where
(\ref{theo}) uses the same basic variables, \cite{T2009}
obtains an enumeration of all NE of $(A,B)$.

Our approach is somewhat similar, with a parameter
$\lambda$ and the equality $x\T a=\lambda$.
However, we consider a different LP which is parameterized
by~$\lambda$ and involves only the payoff matrix $A$ and the
vector $b$ used in $A+B=ab\T$.
That LP, given in (\ref{Plamb}) below, has $x$ as primal and
$y$ as dual variables, whereas in (\ref{theo}) both $x$
and~$y$ are primal with less closely related constraints.
We consider the hyperplane defined by $x\T a=\lambda$ 
\textit{separately} from the parameterized LP.
The intersection of the hyperplane with the
solutions to the parameterized LP defines the 
equilibria of the rank-1 game.
This structural insight can be used both for finding an
exact NE in polynomial time by binary search (see
Section~\ref{s-binsearch}) and for enumerating all
equilibria (see Section~\ref{s-enum}). 
As a topic for further research, it may be interesting if
this approach can be extended to more general linear
multiplicative programs. 

\section{Rank reduction}
\label{s-rankreduce}

The central result of this short section is
Theorem~\ref{t-equiv}.
It states that the set of Nash equilibria of a game of
rank~$r$ is the intersection of a set $\N$ of equilibria of
parameterized games of rank $r-1$ with a suitable
hyperplane.
In subsequent sections, we show how to exploit this property
algorithmically when $r=1$.

The following lemma states the well-known fact that the
equilibria of a bimatrix game are unchanged when
subtracting a separate constant $b_j$ from each
column~$j$ of the row player's payoff matrix.
Call two games \textit{strategically equivalent} if they
have the same Nash equilibria.

\begin{Lemma}
\label{l-add}
If $b\in\reals^n$, then the $m\times n$ game $(A,B)$ is
strategically equivalent to the game $(A-\1 b\T, B)$.
\end{Lemma}

\myproof
This holds by Lemma~\ref{l-nechar}, because the equilibrium
payoff $u$ to player~1 in the game $(A,B)$ changes to $u-b\T
y$ in $(A-\1 b\T, B)$:
Clearly, $Ay\le\1u$ is equivalent to 
$(A-\1b\T)y\le\1(u-b\T y)$, and
$(Ay)_i=u$ is equivalent to 
$((A-\1b\T)y)_i=u-b\T y$.
\myendproof 

\begin{Lemma}
\label{l-plus}
An $m\times n$ bimatrix game of positive rank $r$ can be written
as $(A,C+ab\T)$ for suitable $a\in\reals^m$,
$b\in\reals^n$, and a game $(A,C)$ of rank $r-1$.  
\end{Lemma}

\myproof
An $m\times n$
matrix is of rank at most $r$ if and only if it can be
written as the sum of $r$ rank-1 matrices, that is, as 
$a_1 b_1\T+\cdots+a_r b_r\T$ for
suitable $a_q\in\reals^m$ and $b_q\in\reals^n$ for
$1\le q\le r$. 
This is easily seen by writing the $j$th column of
the matrix as $\sum_{q=1}^r a_q b_{qj}$ and letting
$b_q\T=(b_{q1},\ldots,b_{qn})$
(see also \cite{Wardlaw2005}).
Suppose $(A,B)$ is of rank $r$, with 
$A+B=\sum_{q=1}^{r}a_q b_q\T$ and therefore
$B=-A+\sum_{q=1}^{r}a_q b_q\T$.
Let $C=-A+\sum_{q=1}^{r-1}a_q b_q\T$ and $a=a_r$, $b=b_r$, so that
$B=C+ab\T$; obviously, $A+C$ is of rank $r-1$. 
\myendproof

The following is a simple but central lemma.

\begin{Lemma}
\label{l-equiv}
Let $A,C\in\reals^{m\times n}$,
$x\in X$, $y\in Y$,
$a\in\reals^m$, $b\in\reals^n$, $\lambda\in\reals$.
The following are equivalent:

\myitem{\rm{(a)}}
$(x,y)$ is an equilibrium of $(A,C+a b\T)$,
\myitem{\rm{(b)}}
$(x,y)$ is an equilibrium of $(A,C+\1\lambda b\T)$
and $x\T a=\lambda$,
\myitem{\rm{(c)}}
$(x,y)$ is an equilibrium of
$(A-\1\lambda b\T,C+\1\lambda b\T)$
and $x\T a=\lambda$.
\end{Lemma}

\myproof
The equivalence of (a) and (b) holds because the players
get in both games the same expected payoffs for their pure
strategies:
this is immediate for player~1, and
if $x\T a=\lambda$, then the column payoffs are given by
\begin{equation}
\label{samepay}
x\T(C+a b\T)
=x\T C+ \lambda b\T
=x\T C+ x\T\1\lambda b\T
=x\T(C+\1 \lambda b\T).
\end{equation}
The games in (b) and (c) are strategically equivalent by
Lemma~\ref{l-add}.
\myendproof

Consider a game $(A,B)$ of positive rank $r$ where
$B=C+ab\T$ so that $(A,C)$ is a game of rank $r-1$ according
to Lemma~\ref{l-plus}. Then the game $(A-\1\lambda b\T,C+\1\lambda b\T)$ in 
Lemma~\ref{l-equiv}(c) has the same sum $A+C$ of its
payoff matrices and hence also rank $r-1$, for any choice of
the parameter~$\lambda$.
Let $\N$ be the set of Nash equilibria together with
$\lambda$ of these parameterized games,
\begin{equation}
\label{Nlower}
\N=
\{(\lambda,x,y)\in \reals\times X\times Y \mid
(x,y)\hbox{ is a NE of }
(A-\1\lambda b\T, C+\1\lambda b\T)
\} 
\end{equation}
where by Lemma~\ref{l-equiv}(b)
\begin{equation}
\label{N}
\N=
\{(\lambda,x,y)\in \reals\times X\times Y \mid
(x,y)\hbox{ is a NE of }
(A, C+\1\lambda b\T)
\}.
\end{equation}
These considerations imply the following main result of this
section.

\begin{theorem}
\label{t-equiv}
Given a bimatrix game $(A,C+ab\T)$, its set of
Nash equilibria is exactly the projection on $X\times Y$
of the intersection of $\N$ and the hyperplane $\H$ defined
by 
\begin{equation}
\label{H}
\H=
\{(\lambda,x,y)\in \reals\times\reals^{m}\times\reals^{n}
\mid x\T a=\lambda\}\,.
\end{equation}
\end{theorem}

Theorem~\ref{t-equiv} asserts that for any rank-$r$ game of
the form $(A,C+ab\T)$,
every Nash equilibrium of the game is captured by the set
$\N$ in (\ref{Nlower}) of games of rank $r-1$ which are
parameterized by~$\lambda$, intersected with the hyperplane
$\H$ in (\ref{H}).
Can this rank reduction be leveraged to get an efficient
algorithm to find a Nash equilibrium for a game of arbitrary
constant rank?
As will be discussed in Section~\ref{s-example},
this does not work in general.
However, it does work for rank-1 games. 

\section{Parameterized linear programs}\label{s-paraLP}

Our aim is to describe the equilibria of rank-1 games
$(A,-A+ab\T)$ using the rank reduction of the previous
section. 
For this, we consider the set $\N$ in (\ref{N}) for $C=-A$, 
\begin{equation}
\label{NA}
\N=
\{(\lambda,x,y)\in \reals\times\reals^{m}\times\reals^{n}
\mid
(x,y)\hbox{ is a NE of }
(A,-A+\1\lambda b\T) \} \,,
\end{equation} 
where by (\ref{Nlower}) 
\begin{equation}
\label{NAlower}
\N=
\{(\lambda,x,y)\in \reals\times\reals^{m}\times\reals^{n}
\mid
(x,y)\hbox{ is a NE of }
(A-\1\lambda b\T,-A+\1\lambda b\T) \} \,,
\end{equation} 
which is the set of 
equilibria of zero-sum games parameterized by~$\lambda$.
These correspond to the solutions of a parameterized linear
program (LP).
In this section, we review the structure of such
parameterized LPs with a particular view towards nongeneric
cases and polynomial-time algorithms as studied by
\citet{AdlerM1992}.
In essence, such parameterized LPs
have finitely many special values of the
parameter $\lambda$ called \textit{breakpoints}.
These separate the set $\N$ into a connected sequence of
polyhedral \textit{segments} (which generically are line
segments). 
They are described in Theorem~\ref{t-N} in the next section,
where we will present a polynomial-time algorithm for
finding one equilibrium of a rank-1 game. 
In the subsequent section we present another algorithm for
finding all equilibria.

We assume familiarity with notions of linear programming
such as LP duality and complementary slackness;
see, for example, \cite{Schrijver}. 
The following well-known lemma \cite[p.~286]{Dantzig1963}
states that the equilibria of a zero-sum game are the primal
and dual solutions to an LP.

\begin{Lemma}
\label{l-0sum}
Consider an $m\times n$ zero-sum game $(M,-M)$.
In any equilibrium $(x,y)$ of this game, $y$ is a minmax
strategy of player~2, which is a solution to the LP
with variables $y$ in $\reals^n$ and $u$ in $\reals$:
\begin{equation}
\label{LP}
\maxi_{y,u}u
\quad
\subj
\quad
My+\1u\le\0 ,\quad y\in Y,
\end{equation}
and $x$ is a maxmin strategy of player~1,
which is a solution to the dual LP to $(\ref{LP})$.
For the optimal value of~$u$ in $(\ref{LP})$, the maxmin
payoff to player~1 and minmax cost to player~2 and hence
value of the game is~$-u$. 
\end{Lemma}

\myproof
The dual LP to (\ref{LP}) has variables $x\in\reals^m$ and
$v\in\reals$ and states
\begin{equation}
\label{DLP}
\mini_{x,v}v
\quad \subj \quad
x\T M+v\1\T\ge\0\T,\quad x\in X.
\end{equation}
Both LPs are feasible (with sufficiently small $u$ and large
$v$).
Let $(y,u)$ be an optimal solution to (\ref{LP}) and $(x,v)$
to (\ref{DLP}).
Then $u=v$ by LP duality, and
(\ref{LP}) and (\ref{DLP}) state $My\le\1(-u)$, that is,
player~2 pays no more than $-u$ for any row, and
$x\T M\ge (-v)\1\T$, that is, player~1 gets at least $-v$ in
every column, where $-u=-v$ which is therefore the value of
the game.

With the dual constraints written as $x\T (-M) \le v\1\T$, 
the complementary slackness conditions between the primal
and the dual are exactly the Nash equilibrium conditions
(\ref{xbr}) and (\ref{ybr}) of Lemma~\ref{l-nechar}
(except for the changed sign of~$u$ so that we do not have
to write $x\in X$ in (\ref{DLP}) as $-\1\T x=-1$ and
$x\ge\0$).
Hence, $(x,y)$ is a Nash equilibrium.
\myendproof

Applied to $M=A-\1\lambda b\T$ in (\ref{NAlower}),
the LP~(\ref{LP}) in Lemma~\ref{l-0sum} says:
\begin{equation}
\label{lp0}
\maxi_{y,u}u
\quad \subj \quad
(A - \1\lambda b\T)y+\1u \le\0 \, ,\quad y\in Y.
\end{equation}
In (\ref{lp0}), the matrix $A$ is parameterized.
The substitution $u=\lambda b\T y+t$ 
gives the equivalent LP where only the objective function is
parameterized:
\begin{equation}
\label{lp1}
\maxi_{y,t}\lambda b\T y+t
\quad
\subj
\quad
Ay+\1 t\le\0 ,\quad y\in Y.
\end{equation}
This is a standard parameterized linear programming problem.
We stay close to the notation of \citet{AdlerM1992} who
consider a primal LP with minimization subject to equality
constraints, variables $x$,
and a parameterized right hand side,
of which (\ref{lp1}) is the dual, a maximization
problem subject to inequalities, with variables $y$,
and a parameterized objective function.
We write (\ref{lp1}) as 
\begin{equation}
\label{Dlamb}
D_{\lambda}:\qquad
\maxi_{y,t}\lambda b\T y+t
\quad
\subj
\quad
(y,t)\in D
\end{equation}
with the fixed polyhedron
\begin{equation}
\label{defD}
\begin{array}{rrl}
D=\{\,(y,t)\in\reals^n\times\reals\mid
~~Ay&{}+\1t\le&\0\\
\1\T y&=&1\\
y&\ge&\0~\}\,.\\
\end{array} 
\end{equation}
The LP $D_{\lambda}$ 
is the dual of the following LP $P_{\lambda}$ 
with a parameterized right hand
side, where we use slack variables $s\in\reals^n$ to express
the inequality $A\T x+\1v\ge b\lambda$ as an equality,
in line with \cite{AdlerM1992}: 
\begin{equation}
\label{Plamb}
\arraycolsep.1em
\begin{array}{rrcl}
P_{\lambda}:\qquad
\displaystyle\mini_{x,v,s} v\quad \subj\quad
A\T x&{}+\1 v-s&=&b\lambda\\
\1\T x&&=&1\\
x&,\hfill s&\ge&\0~.\\
\end{array} 
\end{equation} 
For optimal solutions
$(y,t)$ to $D_{\lambda}$
and
$(x,v,s)$ to $P_{\lambda}$
we have $\lambda b\T y+t=v$.
The next lemma (essentially a corollary to
Lemma~\ref{l-equiv} and Lemma~\ref{l-0sum}) states that
$-t$ and $v$ can be interpreted as the player's payoffs for
the games in Lemma~\ref{l-equiv}(a) and~(b),
and asserts that $t,v,s$ are uniquely determined by
$(\lambda,x,y)$ (that is, a point on~$\N$).

\begin{Lemma}
\label{l-pay}
Let $\lambda\in\reals$.
Then $(x,y)$ is an equilibrium of the game
$(A,-A+\1\lambda b\T)$ if and only if $(y,t)$ is an optimal
solution to $D_{\lambda}$ in $(\ref{Dlamb})$ for some $t$
which is uniquely determined by $y$, and $(x,v,s)$ is an
optimal solution to $P_{\lambda}$ in $(\ref{Plamb})$ for
some $v$ and $s$ which are uniquely determined by $\lambda$
and~$x$.
The equilibrium payoffs are $-t$ to player~1 and $v$
to player~2.
If $x\T a=\lambda$, these are also the payoffs in the game
$(A,-A+a b\T)$, and $(x,y)$ is an equilibrium of that game.
\end{Lemma}

\myproof
By Lemma~\ref{l-equiv} with $C=-A$, the game
$(A,-A+ab\T)$ has the same equilibria $(x,y)$ and, by
(\ref{samepay}), payoffs as the game $(A,-A+\1\lambda b\T)$
if $x\T a=\lambda$.
Consider any optimal solutions $(y,t)$ to $D_\lambda$ and
$(x,v,s)$ to $P_\lambda$.
Then $Ay+\1t\le\0$ states for each row~$i$ of $A$ the
inequality $(Ay)_i\le -t$.
Complementary slackness, equivalent to LP optimality, states
that $(Ay)_i=-t$ whenever $x_i>0$.
This is the equilibrium condition in (\ref{xbr})
that states that $x$ is a best response to $y$.
Because it holds for at least one $i$, it uniquely
determines $-t$, which is the equilibrium payoff to player~1
in the above games.

Similarly, the constraint $s=A\T x-b\lambda+\1v$ in
(\ref{Plamb}) means that $s$ is determined by
$(x,\lambda,v)$, and states $s_j=(A\T x-b\lambda)_j+v\ge 0$
for all $j$, or equivalently
$((-A\T +b\lambda\1\T)x)_j\le v$.
Complementary slackness, equivalent to LP optimality, states
that this inequality is tight whenever $y_j>0$.
This is the condition (\ref{ybr}) that states that $y$ is a
best response to $x$ in the game $(A,-A+\1\lambda b\T)$, and
it uniquely determines $v$ as the equilibrium payoff to
player~2.
\myendproof

Primal-dual pairs $P_{\lambda},D_{\lambda}$ of LPs with a
parameter $\lambda$ have been studied since
\cite{GassSaaty1955}.
The next result is well known, which we show following
\cite{jansen1997}.

\begin{Lemma}
\label{l-finite}
For $\lambda\in\reals$, let $\phi(\lambda)$ be the
optimum value of~$P_{\lambda}$ and hence of~$D_{\lambda}$.
Then $\phi:\reals\to\reals$ is the pointwise maximum of a
finite number of affine functions on $\reals$ and therefore
piecewise linear and convex. 
\end{Lemma}

\myproof
The optimum of $D_{\lambda}$ exists for any $\lambda$ and is
taken at a vertex of the polyhedron~$D$ in (\ref{defD}).
Let $V$ be the set of vertices of~$D$, which is finite.
Hence,
\begin{equation}
\label{phi}
\phi(\lambda) = \max \{\lambda(b\T y)+t\mid (y,t)\in V\} 
\end{equation}
where for each of the finitely many $(y,t)$ in $V$ the
function $\lambda\mapsto\lambda(b\T y)+t$ is affine.
Hence, $\phi$ is the pointwise maximum of a finite number of
affine functions as claimed. 
The epigraph of $\phi$ given by $E=\{ (\lambda,\theta)\mid
\theta\ge\phi(\lambda)\}$ is the intersection of the convex
epigraphs of these affine functions, so $E$ is convex and
$\phi$ is a convex function.
\myendproof

By (\ref{phi}), the function $\phi(\lambda)$ is the
``upper envelope'' of the affine functions 
$\lambda\mapsto\lambda(b\T y)+t$ defined by the vertices
$(y,t)$ of~$D$.
A \textit{breakpoint} is any $\lambda^*$ so that
$\phi(\lambda)$
has different left and right derivatives when $\lambda$
approaches $\lambda^*$ from below or above, denoted by
$\phi'_-(\lambda^*)$ and $\phi'_+(\lambda^*)$, respectively.

For any LP $L$, say, let $\OPT(L)$ be the face of the domain
of $L$ where its optimum is attained.
For any $\lambda$ we denote $\OPT(D_\lambda)$ by
$Y(\lambda)$, that is,
\begin{equation}
\label{Ylamb}
Y(\lambda)=\{\,(y,t)\in D\mid \lambda b\T y+t=\phi(\lambda)\}\,.
\end{equation}
Then the left and right derivatives of $\phi$ at $\lambda$
are characterized as follows (obvious from (\ref{phi}), also
Prop.~2.4 of \cite{AdlerM1992}): 
\begin{equation}
\label{LR}
\begin{array}{lll}
\phi'_-(\lambda)&=&
\min\{\,b\T y\mid (y,t)\in Y(\lambda)\}\,,
\\ 
\phi'_+(\lambda)&=&
\max\{\,b\T y\mid (y,t)\in Y(\lambda)\}\,, 
\end{array}
\end{equation}
which are the optima of the two LPs
\begin{equation}
\label{SL}
\renewcommand{\arraystretch}{1.3}
\begin{array}{lll}
\SLmin(\lambda)&:\qquad
\displaystyle\mini_{y,t}
& b\T y \quad\subj\quad (y,t)\in Y(\lambda)\,,\\
\SLmax(\lambda)&:\qquad
\displaystyle\maxi_{y,t}
& b\T y \quad\subj\quad (y,t)\in Y(\lambda)\,.\\
\end{array} 
\end{equation} 
That is, $\lambda^*$ is a breakpoint if and only if
$\phi'_-(\lambda^*)<\phi'_+(\lambda^*)$.
Clearly, in that case there are at least two
vertices $(y,t)$ and $(\hat y,\hat t)$ of~$D$
that define two different
affine functions $\lambda\mapsto\lambda(b\T y)+t$
and $\lambda\mapsto\lambda(b\T \hat y)+\hat t$
that meet at $\lambda=\lambda^*$ to define the maximum
$\phi(\lambda^*)$ in (\ref{phi}).
These are also vertices of $Y(\lambda^*)$, which is then a
higher-dimensional face (such as an edge) of~$D$.
The following central observation
shows that the breakpoints
give all the information about the optimal faces
$Y(\lambda)$ of $D_\lambda$ for any~$\lambda$ between these
breakpoints.

\begin{theorem}
\label{t-endp}
\emph{\cite[Theorem~4.1]{AdlerM1992}~}
Let $\lambda_1,\ldots,\lambda_K$ be
the breakpoints, in increasing order,
for the parameterized LPs
$P_\lambda$ and $D_\lambda$, and let $\lambda_0=-\infty$ and
$\lambda_{K+1}=\infty$.
For $0\le k\le K$, consider any
$\lambda'_k\in(\lambda_k,\lambda_{k+1})$.
Then $Y(\lambda'_k)=\OPT(\SLmax(\lambda_k))$
for $1\le k\le K$, and 
$Y(\lambda'_k)=\OPT(\SLmin(\lambda_{k+1}))$
for $0\le k\le K-1$.
\end{theorem}

For finding the solutions to $P_\lambda$ as a function of
$\lambda$,
the \textit{nondegenerate} case is straightforward, where
$Y(\lambda)$ is a vertex of $D_\lambda$ unless $\lambda$ is
a breakpoint, in which case $Y(\lambda)$ is an edge of
$D_\lambda$.
Then these vertices uniquely describe the pieces of the
piecewise linear function $\phi(\lambda)$, and can be
traversed by a parameterized simplex algorithm \cite{GassSaaty1955}.
An example is shown in the right diagram of Figure~\ref{f1lp}
below with the constraints (\ref{C1}) for $Ay+\1t\le\0$ in~$D$,
with the additional constraints $0\le y_2\le 1$ to represent
$y\in Y$, and objective function $\lambda b\T y+t$ given by
$\lambda(1-2y_2)+t$.
The three linear parts of $\phi(\lambda)$ are
\begin{equation}
\label{phi1}
\phi(\lambda)=\begin{cases}
-\lambda-1&\hbox{for } \lambda\le-\frac12\\
-\frac12& \hbox{for } -\frac12\le\lambda\le\frac12\\
\lambda-1& \hbox{for } \frac12\le\lambda\\
\end{cases}
\end{equation}
which correspond to the optimal vertices $(y_2,t)$ of $D$
given by $(1,-1)$, $(\frac12,-\frac12)$, and $(0,-1)$.
The two breakpoints are $\lambda_1=-\frac12$ and
$\lambda_2=\frac12$ which correspond to the two edges
of~$D$.

In the degenerate case, one typically does not get
polynomial-time algorithms by considering vertices and
corresponding basic solutions to the LP $P_\lambda$ as in a
parameterized simplex algorithm. 
Instead of partitioning the variables of 
$P_\lambda$ into basic and nonbasic variables,
\citet{AdlerM1992} consider ``optimal partitions'';
we use here only the partition part that replaces the
nonbasic variables, which
we denote by $M(\lambda)\cup N(\lambda)$ 
in (\ref{true}) below
(called $N(\lambda)$ in \cite{AdlerM1992}).
This is the set of variables of the dual LP $D_\lambda$ that
may be strictly positive in an optimal solution, which 
represent the ``true inequalities'' of $Y(\lambda)$.

\begin{Definition}
\label{d-truegen}
For some $A,b,C,d$ suppose that the constraints in $x$
\begin{equation}
\label{abcd}
Ax\le b,\qquad Cx=d
\end{equation}
are feasible.
Then any row $i$ of $Ax\le b$ so that $(b-Ax)_i>0$
for some feasible $x$ is called a
\emph{true inequality} of~$(\ref{abcd})$.
\end{Definition}

If there are solutions $x$ and $\hat x$ to (\ref{abcd})
so that $(b-Ax)_i>0$ and $(b-A\hat x)_j>0$ then both
inequalities are true for ${x\frac12+\hat x\frac12}$, so
there is a unique largest set of true inequalities
with some feasible solution where all these strict
inequalities hold simultaneously.
These define the relative interior of the polyhedron defined
by (\ref{abcd}).

Let $A\in\reals^{m\times n}$ and $b\in\reals^n$.
Let $M(\lambda)\cup N(\lambda)$ be the set of true
inequalities of the optimal face $Y(\lambda)$ of $D_\lambda$
in $(\ref{Dlamb})$, that is, 
\begin{equation}
\label{true}
\arraycolsep.1em
\begin{array}{lllrl}
M(\lambda)&=\{\,i\in\{1,\ldots,m\}& \mid&(Ay)_i+t&<0 
~\textrm{ for some }(y,t)\in Y(\lambda)\,\}\,,
\\
N(\lambda)&=\{\,j\in\{1,\ldots,n\}& \mid&y_j&> 0 
~\textrm{ for some }(y,t)\in Y(\lambda)\,\}\,.
\\
\end{array}
\end{equation} 
Any non-true inequality of $Y(\lambda)$ is always tight,
that is, $(Ay)_i+t=0$ if $i\not\in M(\lambda)$ and $y_j=0$
if $j\not\in N(\lambda)$.
It can be shown that for such $i$ and $j$ there are optimal
solutions $(x,v,s)$ to $P_\lambda$ where $x_i>0$ and
$s_j>0$, so these are the true inequalities of
$\OPT(P_\lambda)$.
This is also known as ``strict complementary slackness''
\cite[Section~7.9]{Schrijver}.
Consider the polyhedron $\Primal$ of the constraints for
$P_\lambda$ in (\ref{Plamb}) where $\lambda$ is allowed to
vary, 
\begin{equation}
\label{defQ}
\Primal= \{ (\lambda,x,v,s)
\in\reals\times\reals^m\times\reals\times\reals^n
\mid 
A\T x+\1v-s=b\lambda,~
x\in X,~
s\ge\0\,
\}\,.
\end{equation}
The following lemma considers the face of $\Primal$ defined by the
equations $x_i=0$ for $i\in M(\lambda)$ and $s_j=0$ for
$j\in N(\lambda)$, which are necessary and sufficient for a
feasible solution to $P_\lambda$ to be optimal.
This is immediate from the standard complementary slackness
condition.

\begin{Lemma}
\label{l-Q}
Let $A\in\reals^{m\times n}$ and $b\in\reals^n$.
For $M\subseteq\{1,\ldots,m\}$ and
$N\subseteq\{1,\ldots,n\}$, with 
$x_M=(x_i)_{i\in M}$ and $s_N=(s_j)_{j\in N}$, define
\begin{equation}
\label{QMN}
\Primal(M,N)= \{ (\lambda,x,v,s)\in \Primal \mid x_M=\0,~s_N=\0\,\}\,.
\end{equation}
Then any feasible solution $(x,v,s)$ to $P_\lambda$
is optimal if and only if $(\lambda,x,v,s) \in
\Primal(M(\lambda),N(\lambda))$.
\end{Lemma}

Crucially, according to Theorem~\ref{t-endp}, for any
$\lambda$ in an open interval $(\lambda_{k},\lambda_{k+1})$
(for $0\le k\le K$) the optimal face $Y(\lambda)$ is
constant in~$\lambda$.
Hence,
for all $\lambda\in(\lambda_{k},\lambda_{k+1})$
the true inequalities $(M(\lambda),N(\lambda))$ of
$Y(\lambda)$ are equal to some fixed $(M,N)$,
and for the points $(\lambda,x,v,s)$ in $P(M,N)$ the
value of $\lambda$ can be any real in the \textit{closed}
interval $[\lambda_{k},\lambda_{k+1}]$.
Namely, with the LPs
\begin{equation}
\label{BR}
\renewcommand{\arraystretch}{1.3}
\arraycolsep1pt
\begin{array}{lll}
\BRmax(M,N)&:\quad \displaystyle\maxi_{\lambda,x,v,s}
& \lambda \quad\subj\quad (\lambda,x,v,s)\in \Primal(M,N)\,,\\
\BRmin(M,N)&:\quad\displaystyle \mini_{\lambda,x,v,s}
& \lambda \quad\subj\quad (\lambda,x,v,s)\in \Primal(M,N)\,,\\
\end{array} 
\end{equation} 
the following holds.

\begin{Lemma}
\label{l-BR}
Consider $\lambda_0,\lambda_1,\ldots,\lambda_K,\lambda_{K+1}$ and
$\lambda'_k\in(\lambda_k,\lambda_{k+1})$ for $0\le k\le K$ as
in Theorem~\ref{t-endp}.
Let $M^k=M(\lambda'_k)$ and $N^k=N(\lambda'_k)$
(which do not depend on the choice of $\lambda'_k$).
Then for $1\le k\le K$,
\myitem{\rm{(a)}}
the breakpoint $\lambda_k$ is the
optimum of the LP
$\BRmax(M^{k-1},N^{k-1})$
and of the LP 
$\BRmin(M^{k},N^{k})$;
\myitem{\rm{(b)}}
if $(\lambda,x,v,s)\in \Primal(M(\lambda_{k}),N(\lambda_{k}))$
then $\lambda=\lambda_k$\,.
\end{Lemma}

\myproof
See \cite{AdlerM1992}, p.~171 for (a), and Theorem 3.1(a) and
Lemma~3.1(b) for (b).
\myendproof

Lemma~\ref{l-BR}(a) implies that for any $\lambda$ in the
open interval $(\lambda_{k},\lambda_{k+1})$, for $1\le k\le
K-1$, the endpoints of the closed interval
$[\lambda_{k},\lambda_{k+1}]$ are given by the minimum and
maximum of $\lambda$ for $(\lambda,x,v,s)\in \Primal(M,N)$ where
$M=M(\lambda)$ and $N=N(\lambda)$.
Lemma~\ref{l-BR}(b) and Lemma~\ref{l-Q} imply that if
$\lambda$ is itself a breakpoint, then
$\Primal(M,N)=\{\lambda\}\times\OPT(P_\lambda)$. 

As we will describe in detail in the next section,
Theorem~\ref{t-endp} and Lemma \ref{l-BR} lead to a
description of the set of optimal solutions to $P_\lambda$
and $D_\lambda$ for all $\lambda$ with the help of the
breakpoints $\lambda_1,\ldots,\lambda_K$ in the form of
$2K+1$ polyhedral \textit{segments} (which are lines in the
nondegenerate case).
Any solution $(x,v,s)$ to $P_\lambda$ is optimal if and only
if $(\lambda,x,v,s)$ belongs to $\Primal(M(\lambda),N(\lambda))$,
which is a face of $\Primal$, and  
any solution to $D_\lambda$ is optimal if and only if it
belongs to $Y(\lambda)$, which is a face of~$D$.
For $\lambda$ between two breakpoints, these faces do not
change (but $x$ typically varies with $\lambda$), and their
Cartesian product defines $K+1$ of the segments.
If $\lambda$ is equal to a breakpoint, the set
$\Primal(M(\lambda),N(\lambda))$ is a subset of the two adjoining
faces $\Primal(M(\lambda'),N(\lambda'))$ for $\lambda'$
near~$\lambda$, whereas $Y(\lambda)$ is a superset of the
adjoining faces $Y(\lambda')$, as described in
Theorem~\ref{t-endp}.
This defines the other $K$ segments.
Using this we will give a precise description of the set
$\N$ in Theorem~\ref{t-N} below.

\citet{AdlerM1992} describe how to generate the breakpoints
of $P_\lambda,D_\lambda$ in polynomial time per breakpoint,
with a polynomial-time algorithm applied to the LPs
(\ref{Dlamb}), (\ref{SL}), (\ref{BR}), which we will adapt
to our purpose.
(However, the number of breakpoints may be exponential, see
\cite{Murty1980}.)
The true inequalities in Definition~\ref{d-truegen} can also be
found with an LP, according to the following lemma
(Prop.~4.1 of \cite{AdlerM1992}), due to \cite{FRT1985}; 
for an alternative polynomial-time algorithm
see~\cite{MehrotraYe1993}.

\begin{Lemma}
\label{l-true}
For $A,b,C,d$ and the constraints $(\ref{abcd})$ consider the LP 
\begin{equation}
\label{LPtrue}
\arraycolsep.1em
\begin{array}{rl}
\displaystyle
\maxi_{x,u,\alpha}\1\T u
\quad\subj\quad
Ax{}+u{}-b\alpha&\le\0,
\quad
\\
Cx{}-d\alpha&=0,
\quad
\\
\0\le u&\le\1,
\quad
\\
\alpha&\ge1. 
\end{array}
\end{equation}
Then $(\ref{abcd})$ is feasible if and only if
$(\ref{LPtrue})$ is feasible and bounded, and any optimal
solution $(x^*,u^*,\alpha^*)$ to $(\ref{LPtrue})$ satisfies
$u^*_i=1$ (and $u^*_i=0$ otherwise) if and only if $i$ is a
true inequality of $(\ref{abcd})$.
For such an optimal solution $(x^*,u^*,\alpha^*)$ to $(\ref{LPtrue})$,
$x=x^*(1/\alpha^*)$ is a solution to $(\ref{abcd})$
where $(b-Ax)_i>0$ for all true inequalities~$i$.
\end{Lemma}

\myproof
If the LP $(\ref{LPtrue})$ is feasible then it is also
bounded because $u\le\1$.
Let $I$ be the set of true inequalities of (\ref{abcd}),
that is, $(b-Ax)_i=\varepsilon_i>0$ for $i\in I$ for some
$x$ with $Cx=d$.
Choose $\alpha^*\ge1$ so that $\alpha^*\ge1/\varepsilon_i$
for all $i\in I$.
Then
$(b\alpha^*-A(x\alpha^*))_i=(b-Ax)_i\alpha^*=\varepsilon_i\alpha^*\ge1$
for $i\in I$.
Hence, $x^*=x\alpha^*$ and $u^*$ defined by $u_i^*=1$ if
$i\in I$, and $u_i^*=0$ otherwise, give a feasible solution 
$(x^*,u^*,\alpha^*)$ to the LP (\ref{LPtrue}).
This solution is also optimal because any solution $(\hat
x,\hat u,\hat\alpha)$ to (\ref{LPtrue}) where $\hat u_i>0$
would give a solution $x=\hat x(1/\hat\alpha)$ to
(\ref{abcd}) with $(b-A\hat x)_i>0$ and thus $i\in I$,
so for any feasible solution $(x,u,\alpha)$ to (\ref{LPtrue})
we have $u_i=0$ whenever $i\not\in I$. 
This proves the claim.
\myendproof

\section{Finding one equilibrium of a rank-1 game by binary search}
\label{s-binsearch}

We use the results of the previous section to present a
polynomial-time algorithm for finding one equilibrium of a
rank-1 game $(A,-A+ab\T)$, using binary search for a suitable
value of the parameter $\lambda$ in Theorem~\ref{t-equiv}.
The search maintains a pair of successively closer parameter
values and corresponding equilibria of the game
$(A,-A+\1\lambda b\T)$ that are on opposite sides of the
hyperplane $\H$ in (\ref{H}).
Generically, the set $\N$ in (\ref{NA}) is a piecewise
linear path which has to intersect $\H$ between these two
parameter values.
In general, the \textit{segments} of that ``path'' are
products of certain faces of the polyhedra $D$ in
(\ref{Dlamb}) and $\Primal$ in (\ref{defQ}) described in
Theorem~\ref{t-endp} and Lemma~\ref{l-BR} using the
breakpoints $\lambda_1,\ldots,\lambda_K$ of the LPs
$P_\lambda$ and~$D_\lambda$.

We give a complete description of $\N$ in terms of these
faces of $\Primal$ and $D$, which we project to $\reals\times X$
(for the possible values of $(\lambda,x)$) and $Y$.
Namely, consider 
$\lambda_0,\lambda_1,\ldots,\lambda_K,\lambda_{K+1}$ and
$\lambda'_k\in(\lambda_k,\lambda_{k+1})$ for $0\le k\le K$ as
in Theorem~\ref{t-endp}.
For $0\le k\le K$, define
\begin{equation}
\label{X'}
X'_k=\{(\lambda,x)\mid
(\lambda,x,v,s)\in \Primal(M(\lambda'_{k}),N(\lambda'_{k}))
\,\}\,.
\end{equation}
Note that for any $(\lambda,x,v,s)\in
\Primal(M(\lambda'),N(\lambda'))$ (for any $\lambda'\in\reals$)
the components $v$ and $s$ are uniquely determined by 
$(\lambda,x)$ by Lemma~\ref{l-pay}. 
Similarly, let
\begin{equation}
\label{Y'}
Y'_k=\{y \mid (y,t)\in Y(\lambda'_k)
\,\}
\end{equation}
where again $t$ in $(y,t)$ is uniquely determined by~$y$.
Recall that the choice of 
$\lambda'_k\in(\lambda_k,\lambda_{k+1})$ does not matter for
the definitions of $X'_k$ and $Y'_k$. 
The polyhedra $X'_k\times Y'_k$ for $0\le k\le K$ (which
for $k=0$ and $k=K+1$ are infinite, otherwise bounded)
represent $K+1$ of the segments that constitute $\cal N$
between any two breakpoints $\lambda_k$ and $\lambda_{k+1}$.
They are successively connected by $K$ further segments,
which are polytopes $X_k\times Y_k$ that correspond to the
breakpoints themselves.
These are for $1\le k\le K$ defined by
\begin{equation}
\label{X}
X_k=\{(\lambda,x)\mid
(\lambda,x,v,s)\in \Primal(M(\lambda_{k}),N(\lambda_{k}))
\,\}
\end{equation}
and
\begin{equation}
\label{Y}
Y_k=\{y \mid (y,t)\in Y(\lambda_k)
\,\}\,.
\end{equation}

\begin{theorem}
\label{t-N}
The set $\N$ in $(\ref{NA})$ is given by
\begin{equation}
\label{Nunion}
\N=(X'_0\times Y'_0)~\cup~
\bigcup_{k=1}^K\bigl((X_k\times Y_k)\cup
(X'_k\times Y'_k)\bigr)\,,
\end{equation}
where for $1\le k\le K$ we have
\begin{equation}
\label{cup}
Y_k\supseteq Y'_{k-1} \cup Y'_k
\end{equation}
and
\begin{equation}
\label{cap}
X_k\subseteq X'_{k-1} \cap X'_k~.
\end{equation}
\end{theorem}

\myproof
This follows from Lemma~\ref{l-pay}, Lemma~\ref{l-Q},
and Theorem~\ref{t-endp}.
By Theorem~\ref{t-endp}, $Y(\lambda'_k)$ is the optimal face
of $\SLmax(\lambda_k)$ which is a subset of $Y(\lambda_k)$.
Hence, $Y'_{k}\subseteq Y_k$, and similarly
$Y'_{k-1}\subseteq Y_k$, which implies (\ref{cup}).
In addition, we have $M(\lambda'_k)\subseteq M(\lambda_k)$
and $N(\lambda'_k)\subseteq N(\lambda_k)$ and thus 
$X_k\subseteq X'_{k}$ because of the additional tight
constraints in $\Primal(M(\lambda_k),N(\lambda_k))$. 
Similarly, $X_k\subseteq X'_{k-1}$. 
This shows (\ref{cap}). 
\myendproof 

The preceding characterization of $\N$ is used in the
following lemma.

\begin{Lemma}
\label{l-cont}
Let $\llam\leqslant\ulam$ and $\lx,\ux\in X$ and $\ly,\uy\in Y$ so
that
for $\N$ in $(\ref{NA})$ 
\begin{equation}
\label{bounds}
(\llam,\lx,\ly)\in\N\,,
\qquad
\llam\leqslant\lx\T a\,,
\qquad
(\ulam,\ux,\uy)\in\N\,,
\qquad
\ux\T a\leqslant\ulam\,.
\end{equation}
Then $x\T a=\lambda$ for some $(\lambda,x,y)\in\N$ with
$\lambda\in[\llam,\ulam]$.
\end{Lemma}

\myproof
Consider the largest $\lambda^*$ so that
$\lambda^*\in[\llam,\ulam]$ and there are $x^*,y^*$ with
$(\lambda^*,x^*,y^*)\in\N$ and $\lambda^*\le{x^*}\T a$,
which exists since $\llam$ fulfills this property and $\N$
is closed by Theorem~\ref{t-N}.

If $\lambda^*=\ulam$ then both $(\lambda^*,\ux)$ and
$(\lambda^*,x^*)$ belong
to the same set $X_k$ or $X'_k$ which is convex, where since
$\ux\T a\le\lambda^*$ and $\lambda^*\le{x^*}\T a$
we have $x\T a=\lambda^*$ for a suitable convex
combination $x$ of $\ux$ and $x^*$,
and $(\lambda^*,x,y^*)\in\N$, as claimed.

Hence, we can assume $\lambda^*<\ulam$.
Suppose $\lambda^*$ is a breakpoint $\lambda_k$, so that
$(\lambda^*,x^*)\in X_k$.
Consider $\lambda'\in(\lambda_k,\min\{\lambda_{k+1},\ulam\})$ 
and $(\lambda',x',y')\in X'_k\times Y'_k$
where $\lambda'>{x'}\T a$ by maximality of $\lambda^*$.
By (\ref{cap}), we have $(\lambda^*,x^*)\in X'_k$
and hence $(\lambda^*,x^*,y')\in X'_k\times Y'_k$.
Because $\lambda^*\le {x^*}\T a$ and $\lambda'>{x'}\T a$,
a suitable convex combination $(\lambda,x,y')$ of
$(\lambda^*,x^*,y')$ and $(\lambda',x',y')$ belongs to $\N$
and fulfills $\lambda=x\T a$ as claimed (in fact,
$(\lambda,x,y')=(\lambda^*,x^*,y')$ does by maximality of
$\lambda^*$).
If $\lambda^*$ is not a breakpoint, we directly have
$(\lambda^*,x^*,y^*)\in X'_k\times Y'_k$ for some $k$ and
can choose $(\lambda',x',y^*)\in X'_k\times Y'_k$
with $\lambda^*<\lambda'\le\ulam$ and apply the same
argument.
\myendproof

The binary search algorithm will maintain (\ref{bounds}) as
an \textit{invariant} while halving the length of the
interval $[\llam,\ulam]$ in each iteration.

Lemma~\ref{l-cont} ensures that the interval contains some
$\lambda$ with $(\lambda,x,y)\in\N$ and $x\T a=\lambda$
(which is not true when applied to games of higher rank, as
shown in the example in Figure~\ref{f2} below).
Let $\lambda'=(\llam+\ulam)/2$ and let $x'$ be the strategy 
of player~1 in an
equilibrium $(x',y')$ of the game ${(A,-A+\1\lambda'b\T)}$,
which is found as a solution $(x',v',s')$ to $P_{\lambda'}$.
If $\lambda'\le{x'}\T a$, it is natural to proceed with
$\llam$ set to $\lambda'$ (written as
$\llam\leftarrow\lambda'$), otherwise with
$\ulam\leftarrow\lambda'$.
The binary search should terminate once $\llam$ and $\ulam$
are in the same interval $[\lambda_k,\lambda_{k+1}]$ between
two breakpoints, with the desired equilibrium found in
$(X'_k\times Y'_k)\cap\H$.

However, this straightforward approach has the following
problems:
{\myitem{(i)}
the search may converge to an equilibrium $(x,y)$ with $x\T
a=\lambda$ where $\lambda$ is a breakpoint $\lambda_k$, so
that $\llam$ and $\ulam$ are always in different intervals
$(\lambda_{k-1},\lambda_k]$ and $[\lambda_k,\lambda_{k+1})$
and the described termination condition fails;
\myitem{(ii)}
the number of digits to describe $\llam$ and $\ulam$ may
pile~up, which slows down solving $P_{\lambda'}$.}

\begin{figure}[htb]
\[
\includegraphics[height=80mm]{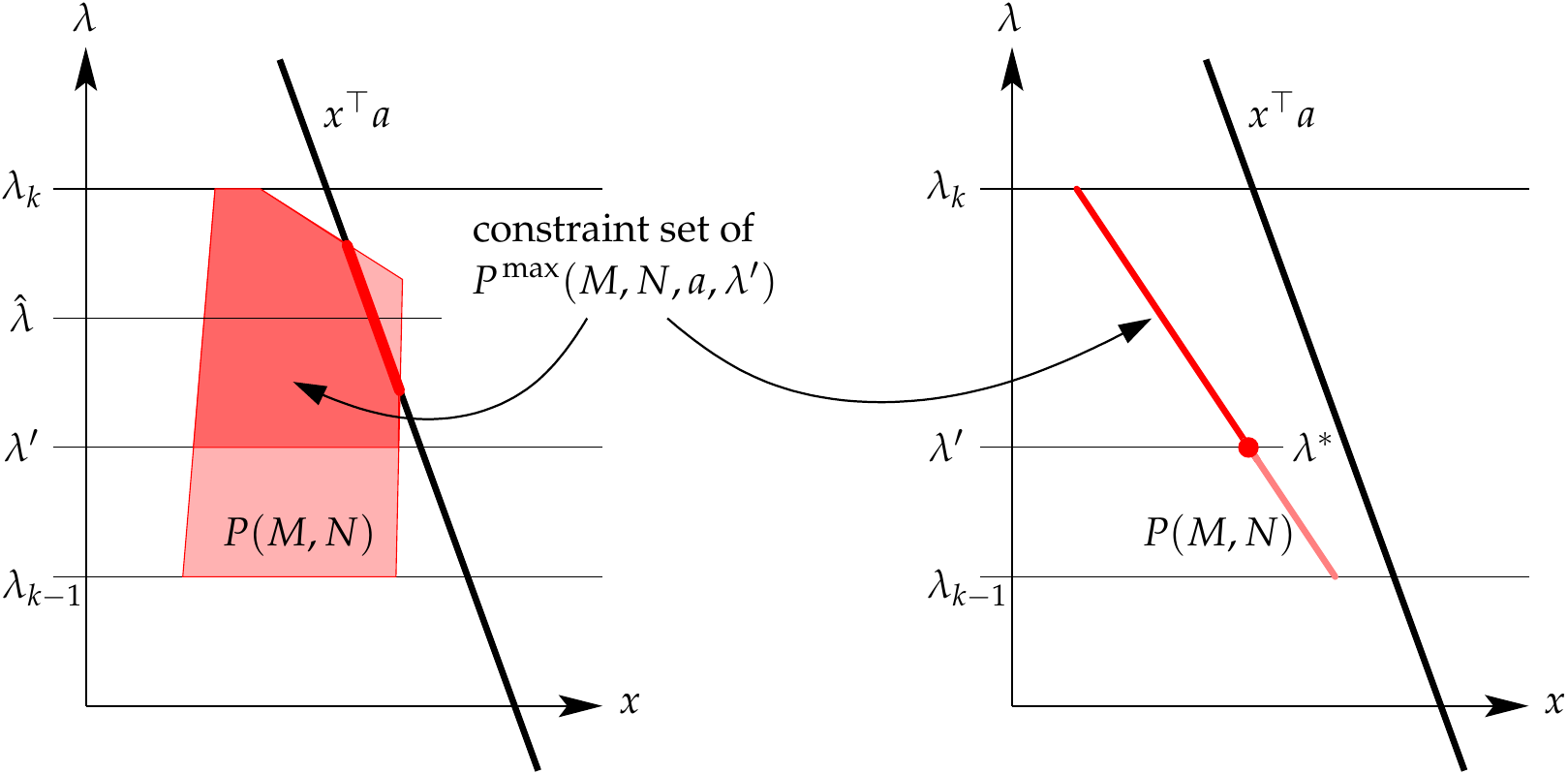}%
\]
\caption{%
(Color online)
Illustration of $\Qmax(M,N,a,\lambda')$ in (\ref{Q}) 
for $\lambda'\in(\lambda_{k-1},\lambda_k)$,
with $M=M(\lambda')$, $N=N(\lambda')$, 
and $P(M,N)$ as a polytope (left) or line segment (right)
}
\label{fqmax}
\end{figure}

\noindent
We address these problems as follows.
First, we identify with $M=M(\lambda')$, $N=N(\lambda')$ the
face $\Primal(M,N)$ of~$\Primal$ that contains
$(\lambda',x',v',s')$.
We then check if that face contains some $(\lambda,x,v,s)$
with ${x\T a=\lambda}$.
Depending on whether 
$\lambda'\le {x'}\T a$
or
${x'}\T a\le \lambda'$,
this is achieved by one of 
the following variations of the LPs in (\ref{BR}) (these
variations will also be used for the enumeration of all
equilibria in Section~\ref{s-enum}): 
\begin{equation}
\label{Q}
\arraycolsep1pt
\renewcommand{\arraystretch}{1.3}
\begin{array}{llll}
\Qmax(M,N,a,\lambda'):&\quad
\displaystyle\maxi_{\lambda,x,v,s} & \lambda-x\T a
\\
&\quad\subj\quad&(\lambda,x,v,s)\in \Primal(M,N)\,,
\\&&
x\T a\ge\lambda\ge\lambda'\,,
\\[1ex]
\Qmin(M,N,a,\lambda'):&\quad
\displaystyle\mini_{\lambda,x,v,s} & \lambda-x\T
a
\\
&\quad\subj\quad&(\lambda,x,v,s)\in \Primal(M,N)\,,
\\&&
x\T a\le\lambda\le\lambda'\,.
\\
\end{array} 
\end{equation} 
Figure~\ref{fqmax} illustrates $\Qmax(M,N,a,\lambda')$ where
$\lambda'<{x'}\T a$, and
$\lambda'$ is between two breakpoints $\lambda_{k-1}$
and $\lambda'_k$ (but $\lambda'$ could also be a breakpoint
itself), so that $P(M,N)$ is projected to $X'_{k-1}$.
Here the optimal solution $x'$ to $P_{\lambda'}$ is not
unique, but always fulfills $\lambda'<{x'}\T a$.
Moreover, $X'_{k-1}\times Y'_{k-1}$ and $H$ intersect. 
In the left diagram in Figure~\ref{fqmax}, $P(M,N)$ is not
just a line segment but a higher-dimensional polytope.
It contains some $(\lambda,x,v,s)$ and
$(\lambda,\hat x,\hat v,\hat s)$
with $x\T a<\lambda< \hat x\T a$, for example
for $\lambda=\hat\lambda$, but not for $\lambda=\lambda'$
nor $\lambda=\lambda_k$. 
In the right diagram of Figure~\ref{fqmax}, we always have
$\lambda<x\T a$, and $\Qmax(M,N,a,\lambda')$ attains its
optimum $\STAR\lambda$ at $\lambda'$ because for the
corresponding $(\STAR x,\STAR\lambda)$, shown as a dot,
$\STAR\lambda-\STAR x\T a$ is least negative.
Here, the solution $\STAR\lambda=\lambda_k$ would be more
useful for proceeding because it is the next breakpoint.
We will introduce an extra computation step to achieve this,
as we discuss further below. 

The next lemma states that the appropriate LP in (\ref{Q}) 
identifies if there is an equilibrium $(x,y)$ of
the game ${(A,-A+\1\lambda b\T)}$ with $x\T a=\lambda$ for
some $\lambda$ between $\lambda'$ and the next breakpoint
$\lambda_k$\,.

\begin{Lemma}
\label{l-QQ}
Let $\lambda_k$ be a breakpoint of $P_\lambda$ and
$D_\lambda$ as in Theorem~\ref{t-endp}, $1\le k\le K$.
Let $\lambda'\in\reals$, let $(x',v',s')$ be an optimal solution
to $P_{\lambda'}$, and let $(M,N)=(M(\lambda'),N(\lambda'))$
as in $(\ref{true})$.
\myitem{\rm(a)}
Suppose $\lambda'\in(\lambda_{k-1},\lambda_k]$ and
$\lambda'\le{x'}\T a$.
Let $(\STAR\lambda,\STAR x,\STAR v,\STAR s)$
be an optimal solution to $\Qmax(M,N,a,\lambda')$. 
Then $\STAR\lambda\in[\lambda',\lambda_k]$, and
the game $(A,-A+\1\lambda b\T)$ has an equilibrium
$(x,y)$ with $x\T a=\lambda$ for some
$\lambda\in[\lambda',\lambda_k]$
if and only if this holds for $\lambda=\STAR\lambda$
and $x=\STAR x$.
\myitem{\rm(b)}
Suppose $\lambda'\in[\lambda_k,\lambda_{k+1})$ and
$\lambda'\ge{x'}\T a$.
Let $(\STAR\lambda,\STAR x,\STAR v,\STAR s)$
be an optimal solution to $\Qmin(M,N,a,\lambda')$. 
Then $\STAR\lambda\in[\lambda_k,\lambda']$, and
the game $(A,-A+\1\lambda b\T)$ has an equilibrium
$(x,y)$ with $x\T a=\lambda$ for some
$\lambda\in[\lambda_k,\lambda']$
if and only if this holds for $\lambda=\STAR\lambda$
and $x=\STAR x$. 
\end{Lemma}

\myproof
We prove (a), where (b) is entirely analogous.
By Lemma~\ref{l-Q}, $(\lambda',x',v',s')$ is feasible for
$\Qmax(M,N,a,\lambda')$. 
Clearly $\lambda'\le\STAR\lambda$, and
Lemma \ref{l-BR} implies $\STAR\lambda\le\lambda_k$.
Because $\lambda\le x\T a$ for any feasible solution
$(\lambda,x,v,s)$, the objective function $\lambda-x\T a$ is
nonpositive, and zero and hence optimal if and only if
$\lambda=x\T a$, in which case $x$ is part of the described
equilibrium $(x,y)$.
\myendproof

\begin{figure}[hbt]
\hrule\vskip2ex
\noindent
\includegraphics[height=138mm]{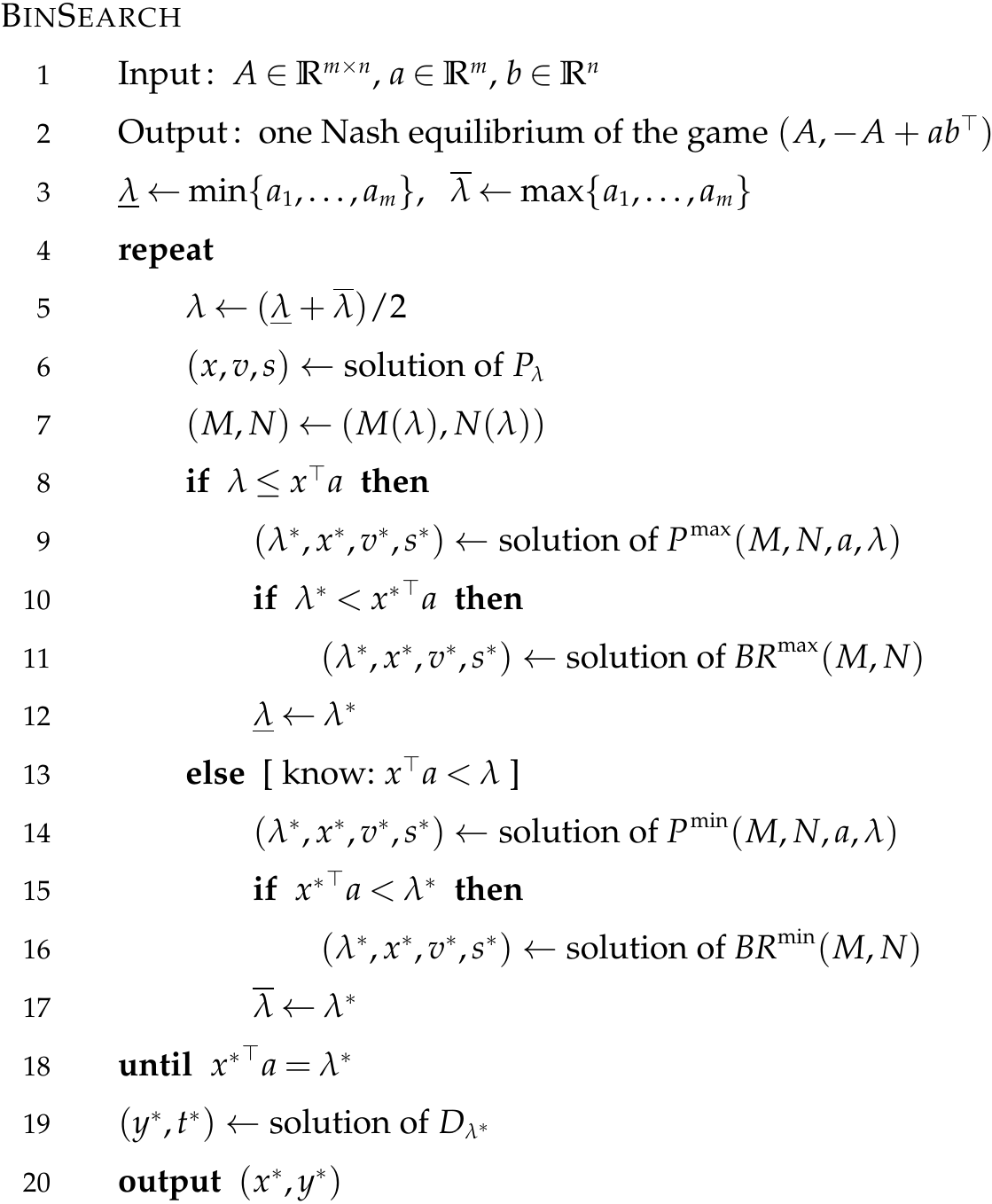}
\vskip2ex\hrule
\vskip.3ex
\caption{%
The \BINSEARCH{} algorithm for finding
one Nash equilibrium of a rank-1 game $(A,-A+ab\T)$
}
\label{fbin}
\end{figure}

We now describe the \BINSEARCH{} algorithm in
Figure~\ref{fbin}, where we will return to the LPs in (\ref{Q}).
The conditions $x\T a=\lambda$ and $x\in X$ mean that
$\lambda$ is a convex combination of the components
$a_1,\ldots,a_m$ of $a$, so that we can initialize
$\llam$ and $\ulam$ as their minimum and maximum
in line~3 of the algorithm.
The main loop of the algorithm is between lines 4 and~18.
The candidate value for $\lambda$ (called $\lambda'$ in the
above explanations) is the midpoint between $\llam$ and
$\ulam$ in line~5.
Line~6 computes some optimal solution
$(x,v,s)$ of the LP~$P_\lambda$ in (\ref{Plamb}),
where the dual LP $D_\lambda$ in (\ref{Dlamb}) is
typically solved alongside~$P_\lambda$.
The optimum $\phi(\lambda)$ of
$P_\lambda$ and $D_\lambda$ determines the optimal face
$Y(\lambda)$ of $D_\lambda$ in (\ref{Ylamb}).
The true inequalities $M,N$ of $Y(\lambda)$ in line~7
are determined according to (\ref{true}), for example with
the help of the LP in Lemma~\ref{l-true}.

Lines~8 to~12, and symmetrically
13 to~17, use the LPs in (\ref{Q}).
In order to match the notation in the discussion before
Lemma~\ref{l-QQ}, let $\lambda'=\lambda$.
Consider the case ${\lambda'\le x\T a}$, handled in
lines~8 to~12. 
Line~9 invokes the LP~ $\Qmax(M,N,a,\lambda')$.
By Lemma~\ref{l-QQ}, the optimum
$(\STAR\lambda,\STAR x, \STAR v, \STAR s)$
to this LP will find the desired equilibrium with
$\STAR\lambda=\STAR x\T a$ if there is one for some
$\STAR\lambda$ up to the next breakpoint $\lambda_k$, that
is, for $\STAR\lambda\in[\lambda',\lambda_k]$.
Suppose this is not the case, that is, 
$\STAR\lambda<\STAR x\T a$ and the optimum 
$\STAR\lambda-\STAR x\T a$ of $\Qmax(M,N,a,\lambda')$ is
negative.
By Lemma~\ref{l-QQ}, in this case the next breakpoint
$\lambda_k$ does \textit{not} define an equilibrium, so that
problem (i) above does not occur.
However, as shown in the right diagram in
Figure~\ref{fqmax}, this may result in
$\STAR\lambda=\lambda'$\,.
We could simply continue with $\llam\leftarrow\STAR\lambda$ as
in line 12, but if $\STAR\lambda=\lambda'$ this
increases the description size of $\llam$ which we would like
to keep bounded to avoid problem (ii) (the description size
of $\lambda$ probably increases only by one bit per main
iteration, but it is useful to keep it independent of the
number of iterations both for the computation and for the
analysis). 
In line 10, the condition
$\STAR\lambda<\STAR x\T a$ recognizes that the current
segment of~$\N$ contains no equilibrium, and then
$\BRmax(M,N)$ in line 11 computes $\STAR\lambda$
as the next breakpoint $\lambda_k$ according to
Lemma~\ref{l-BR}(a);
the LP in line 11 can be solved by starting from
the current solution to $\Qmax(M,N,a,\lambda')$.
The left diagram in Figure~\ref{fqmax} shows that we cannot
simply replace the objective function $\lambda-x\T a$ of
$\Qmax(M,N,a,\lambda')$ by $\lambda$: 
While this would compute the next breakpoint $\lambda_k$\,, it
may overlook that the current segment of $\N$ defined by
$P(M,N)$ intersects the hyperplane~$\H$; this could
possibly miss the equilibrium altogether, for example if
$\ulam=\hat\lambda$ as shown in the diagram (in particular
if $\ulam$ still has its initial value, which is not checked
in the algorithm as to whether it produces an equilibrium).

In summary: lines~8 to~11 
find $\STAR\lambda$ and $\STAR x$ so that either (a)
$\STAR x\T a=\STAR\lambda$, or (b)
$\STAR\lambda<\STAR x\T a$ and $\STAR\lambda$ is a
breakpoint and
$(\llam+\ulam)/2=\lambda\le\STAR\lambda<\ulam$,
which implies $\ulam-\STAR\lambda\le (\ulam-\llam)/2$\,.
The next value of $\llam$ is set to $\STAR\lambda$ in
line~12.
In case (a), the loop terminates in line~18.
In case (b), the loop continues, and in the next
iteration the difference $\ulam-\llam$ has shrunk by at
least one half.
The analogous statements hold for lines
13--17. 
The following theorem states the correctness and polynomial
running time of the algorithm.

\begin{theorem}
\label{t-binsearch}
Algorithm \BINSEARCH{} finds one equilibrium of the
rank-1 game $(A,-A+ab\T)$.
Assume that the entries of $A,a,b$ are rational numbers with
combined bit length~$L$, and that LPs are solved with
polynomial-time solvers that return extreme LP solutions
obtained from linear equations derived from $A,a,b$.
Then \BINSEARCH{} runs in polynomial time in~$L$.
\end{theorem}

\myproof
During the main loop, the invariant (\ref{bounds}) is
preserved, and the length of the interval $[\llam,\ulam]$
shrinks by at least a factor of two per iteration.
By Lemma~\ref{l-cont}, a solution $(\lambda,x,y)\in\N$ with
$x\T a=\lambda$ and $\lambda\in[\llam,\ulam]$ is guaranteed
to exist. 
The termination condition $\STAR x\T a= \STAR\lambda$ in
line~18 holds once $\lambda$ reaches a segment of
$\N$ that intersects $\H$, which is identified with one of
the LPs in line~9 or~14 by
Lemma~\ref{l-QQ}.
Because the length of the search interval $[\llam,\ulam]$
shrinks by at least half in each iteration, the search
interval eventually contains at most one
breakpoint~$\lambda_k$.
If there is no breakpoint in $[\llam,\ulam]$,
then
$(M(\llam),N(\llam))=(M(\ulam),N(\ulam))=(M(\lambda),N(\lambda))$
for $\lambda=(\llam+\ulam)/2$.
Hence, a solution $(\STAR\lambda,\STAR x,\STAR v,\STAR s)$
to $\Qmax(M(\lambda),N(\lambda),a,\lambda)$ or to
$\Qmin(M(\lambda),N(\lambda),a,\lambda)$ determines an
equilibrium $(\STAR x,\STAR y)$ to $(A,-A+ab\T)$ by
Lemma~\ref{l-QQ} and Lemma~\ref{l-equiv}.
This holds also if there is a single breakpoint $\lambda_k$
in $[\llam,\ulam]$.
Hence, as claimed, the algorithm computes an equilibrium 
$(\STAR x,\STAR y)$ of $(A,-A+ab\T)$.

The number of overall iterations is polynomial for the
following reason.
Any breakpoint $\lambda$ is part of a vertex
$(\lambda,x,v,s)$ of $\Primal$ by Lemma~\ref{l-BR}(a).
This vertex is a solution to a linear system of equations
where each component (such as $\lambda$) is a fraction
with an integer determinant obtained from $A,b$ in the
denominator (which has a polynomial of bits), and distinct
fractions for different breakpoints~$\lambda$.
Hence, any two breakpoints have minimum distance
$1/2^{p(L)}$ for some polynomial~$p$
(see also \citep[Section 10.2]{Schrijver}).
Therefore, there will be at most $O(p(L))$ binary
search iterations until the search interval contains at most
one breakpoint and the search terminates.

Each iteration of the algorithm solves three or four LPs.
The first is $P_\lambda$ in line~6.
Using the optimum $\phi(\lambda)$ of that LP, in
line~7 the true inequalities in (\ref{true}) of
$Y(\lambda)$ in (\ref{Ylamb}) are found with another LP as
in Lemma~\ref{l-true}.
The third LP is either $\Qmax(M,N,a,\lambda)$ in
line~9 or $\Qmin(M,N,a,\lambda)$ in
line~14.
The fourth LP is either $\BRmax(M,N)$ or $\BRmin(M,N)$ in
line 11 or~16, respectively (which
just relaxes the extra constraints of the previous LP in
(\ref{Q}) and has a different objective function).
In all cases, the output $\STAR\lambda$ is described in
terms of $A,a,b$ and found in polynomial time in the bit
size~$L$, and $\STAR\lambda$ itself has polynomial bit size 
\citep[Corollary 10.2a(iii)]{Schrijver}.
In the next iteration, $\STAR\lambda$ determines with the
constant arithmetic expression in line~{5} the
next parameter $\lambda$ for $P_{\lambda}$ in
line~6 and for $(M,N)$ in line~7 so that the
bit size of $\lambda$ remains polynomial in~$L$.
Hence, each main iteration takes polynomial time, and the
overall running time is polynomial.
\myendproof

In practice, as observed in \cite[Section~5]{AdlerM1992}, in
the nondegenerate case the segments of $\N$ are line
segments. 
Then the LP in line 9 or~14 is solved
starting from the current solution to $P_\lambda$ in
line~6 with a single pivot, and similarly the
next LP in line 11 or~16.

\section{Enumerating all equilibria of a rank-1 game}
\label{s-enum}

In this section, we show how to obtain a complete
description of all Nash equilibria of a rank-1 game
with the help of Theorem~\ref{t-equiv} and
Theorem~\ref{t-N}.

A degenerate bimatrix game may have infinite sets of Nash
equilibria.
They can be described via \textit{maximal Nash subsets}
\citep{jansen1981maximal}, called ``sub-solutions'' by
\citet{Nash1951}.
A Nash subset for $(A,B)$ is a nonempty product set $S\times
T$ where $S\subseteq X$ and $T\subseteq Y$ so that every
$(x,y)$ in $S\times T$ is an equilibrium of $(A,B)$; in
other words, any two equilibrium strategies $x\in S$ and
$y\in T$ are ``exchangeable''.
Using the ``best response polyhedra'' $\HP$ and $\HQ$
in~(\ref{HPQ}), it can be shown that any maximal Nash subset
$S\times T$ is a polytope, with $S$ as a suitable face of
$\HP$ projected to $X$, and $T$ as a suitable face of $\HQ$
projected to $Y$ \citep{ARSvS}.
These faces are defined by converting some inequalities in
(\ref{HPQ}) to equations, which have to fulfill the 
equilibrium conditions (\ref{xbr}) and~(\ref{ybr}).
The usual output for ``enumerating'' all equilibria consists
of listing all maximal Nash subsets $S\times T$ via the
vertices of $S$ and $T$.
These are vertices of $\HP$ and $\HQ$, respectively
(projected to $X$ and $Y$) that define the ``extreme'' Nash
equilibria of $(A,B)$, with maximal Nash subsets obtained as
maximally exchangeable sets 
of such vertices \citep[Prop.~4]{ARSvS}.
Maximal Nash subsets may intersect, in which case their
vertex sets intersect.
In a nondegenerate game, all maximal Nash subsets are
singletons.

For a rank-1 game $(A,-A+ab\T)$, its set of Nash equilibria
is $\N\cap\H$ projected to $X\times Y$
by Theorem~\ref{t-equiv}, with $\N$ in (\ref{NA}) and $\H$
in~(\ref{H}).
By (\ref{Nunion}), $\N$ is the union of polyhedra, whose
nonempty intersections with $\H$ give almost directly the
maximal Nash subsets.

\begin{theorem}
\label{t-maxnash}
Let $(A,-A+ab\T)$ be a rank-1 bimatrix game, and let 
$\lambda_0,\lambda_1,$ $\ldots,\lambda_K,\lambda_{K+1}$ and
$\lambda'_k\in(\lambda_k,\lambda_{k+1})$ for $0\le k\le K$ as
in Theorem~\ref{t-endp}.
With $(\ref{X'})$, $(\ref{Y'})$, $(\ref{X})$, $(\ref{Y})$, 
let
\begin{equation}
\label{Sr}
\begin{array}{rcll}
S_k&=&\{\,x\mid (\lambda,x)\in X_k\,,~x\T a=\lambda\,\}
&\qquad(1\le k\le K)
, \\
L_k&=&\{\,\lambda\mid (\lambda,x)\in X_k\,,~x\T a=\lambda\,\}
&\qquad(1\le k\le K)
, \\
S'_k&=&\{\,x\mid (\lambda,x)\in X'_k\,,~x\T a=\lambda\,\}
&\qquad(0\le k\le K)
, \\
L'_k&=&\{\,\lambda\mid (\lambda,x)\in X'_k\,,~x\T a=\lambda\,\}
&\qquad(0\le k\le K)
.
\end{array}
\end{equation}
Then the maximal Nash subsets of $(A,-A+ab\T)$ are the sets
$S_k\times Y_k$ if $S_k\ne\emptyset$, and
$S'_k\times Y'_k$ if $S'_k\ne\emptyset$ and $L'_k$ is not
equal to $\{\lambda_k\}$ or $\{\lambda_{k+1}\}$. 
\end{theorem}

\myproof
Each set
$S_k$ is the projection of $(X_k\times Y_k)\cap\H$ on $X$,
and
$S'_k$ is the projection of $(X'_k\times Y'_k)\cap\H$ on
$X$,
with $L_k$ and $L'_k$ containing the corresponding set of
$\lambda$'s.
Hence, by Theorem~\ref{t-N}, if $S_k\ne\emptyset$ then
$S_k\times Y_k$ is a Nash subset, and if
$S'_k\ne\emptyset$ then $S'_k\times Y'_k$
is a Nash subset, and the union of these is the 
set of all equilibria which is the projection of
$\N\cap\H$ on $X\times Y$ by Theorem~\ref{t-equiv}.
The only question is which of these Nash subsets are
inclusion-maximal.
By Corollary~3.2 of \cite{AdlerM1992}, 
$Y_k\cap Y_{k+1}=Y'_k$ where
$Y_k$ and $Y_{k+1}$ contain $Y'_k$ properly,
$Y_k\cap Y_\ell=\emptyset$ whenever $|k-\ell|\ge 2$, and
$Y'_k\cap Y'_\ell=\emptyset$ whenever $k\ne\ell$, and
Lemma~\ref{l-BR} implies
$L_k=\{\lambda_k\}=L_{k-1}\cap L_k$.
So the only possible inclusions are that
$S'_k\times Y'_k$ is a subset of 
$S_k\times Y_k$ or of
$S_{k+1}\times Y_{k+1}$.
Suppose $x\in S'_k$, that is, $(\lambda,x)\in X'_k$ and
$x\T a=\lambda$.
If this implies $\lambda=\lambda_k$ then
$L'_k=\{\lambda_k\}$.
By Lemma~\ref{l-Q}, this means $x$ is part of an optimal
solution $(x,v,s)$ to $P_{\lambda_k}$ and hence
$x\in S_k$, which shows the proper inclusion
$S'_k\times Y'_k\subset S_k\times Y_k$ because
$Y'_k\subset Y_k$.
Similarly, $L'_k=\{\lambda_{k+1}\}$ implies
$S'_k\times Y'_k\subset S_k\times Y_{k+1}$.
These are the only possible inclusions
because if $x\in S'_k$ with $(\lambda,x)\in X'_k$
so that $x\T a=\lambda\not\in\{\lambda_k,\lambda_{k+1}\}$ 
we clearly cannot have $x\in S_k$, say, where $x\T
a=\lambda_k$.

This proves the theorem.
We also note that the described sets $S_k$ and $S'_k$ are
defined in terms of the game $(A,-A+ab\T)$ independently of
the parameter $\lambda$.
Namely, the condition $x\T a=a\T x=\lambda$ implies that 
the polyhedron $\HP$ in (\ref{HPQ})
for $B=-A+ab\T$
is given by
\begin{equation}
\label{HPface}
\begin{array}{rcl}
\HP&=&
\{(x,v)\in X\times\reals\mid
(-A+ab\T)\T x\le\1v\,\}
\\
&=&
\{(x,v)\in X\times\reals\mid
-A\T x+b\lambda\le\1v\,\}\,,
\\
\end{array}
\end{equation}
so $S_k$ and $S'_k$ are projections of certain faces of $\HP$.
\myendproof

A suitable algorithm that enumerates all Nash equilibria can
be adapted from the algorithm by \citet[p.~173]{AdlerM1992}
that proceeds from breakpoint to breakpoint using
Theorem~\ref{t-endp}.
The corresponding segments of $\N$ can then be checked for
nonempty intersections with $\H$, which are then output as
maximal Nash subsets if they meet the conditions of
Theorem~\ref{t-maxnash}.

We give an outline of this algorithm.
Suppose $\lambda$ is equal to a breakpoint $\lambda_k$.
Then $Y_k$ in (\ref{Y}) is the projection of
$Y(\lambda_k)=\OPT(D_{\lambda_k})$,
and $X_k$ in (\ref{X}) is the projection of 
$\OPT(P_{\lambda_k})$ by Lemma~\ref{l-BR}(b) and
Lemma~\ref{l-Q}.
If $(X_k\times Y)\cap\H$ is not empty, its projection to
$X\times Y$ is a maximal Nash subset $S_k\times Y_k$.
Start from some $(\lambda,x)\in X_k$.
If $\lambda=x\T a$ then $x\in S_k$, which is a suitable
starting point for the vertex enumeration of the
polytope~$S_k$, for example with the program \textit{lrs}
\citep{avis2000}.
If $\lambda<x\T a$ or $\lambda>x\T a$ then the condition
$(X_k\times Y)\cap\H\ne\emptyset$ is checked with one of the
LPs in (\ref{Q}) by Lemma~\ref{l-QQ} which then have optimal
value zero, with optimum
$(\STAR\lambda,\STAR x,\STAR v,\STAR s)$;
then $\STAR\lambda=\STAR x\T a$, and $\STAR x\in S_k$
is a new starting point to enumerate the vertices of~$S_k$.

The next segment to be tested for its intersection with $\H$
is $X'_k\times Y'_k$ in (\ref{X'}) and (\ref{Y'}).
For that purpose it is not necessary to find some
$\lambda'\in(\lambda_k,\lambda_{k+1})$, because
$Y(\lambda')=\OPT(\SLmax(\lambda_k))$ by
Theorem~\ref{t-endp}, and the true inequalities $M\cup N$ of
that face are found by Lemma~\ref{l-true}, so that one
obtains $X'_k$ as the projection of $\Primal(M,N)$.
Moreover, we have $x\in X_k\subseteq X'_k$.
If $\lambda=x\T a$ then $x$ is also a starting point for the
enumeration of the vertices of $S'_k$, which gives the Nash
subset $S'_k\times Y'_k$ (which is, however, not maximal if
$S'_k\subseteq S_k$, see Theorem~\ref{t-maxnash}).
If $\lambda<x\T a$ then we solve $\Qmax(M,N,a,\lambda_k)$
in (\ref{Q}) to find out if $\H$ intersects the current
segment $X'_k\times Y'_k$, and similarly
$\Qmin(M,N,a,\lambda_k)$ if $\lambda>x\T a$.
Finally, the next breakpoint $\lambda_{k+1}$ is found
as the solution to $\BRmax(M,N)$ in (\ref{BR}) by
Lemma~\ref{l-BR}(a).

For initialization and termination of this algorithm, we use
that the possible values of $\lambda$ can be restricted to
$[\la,\ua]$ with $\la$ and $\ua$ as minimum and maximum of
$\{a_1,\ldots,a_m\}$.
The initialization is $\lambda=\la$, which is decided to be
a breakpoint or not as described after~(\ref{SL}).
The constraint $\lambda\le\ua$ is added to the step of
finding the next breakpoint, which terminates the algorithm
when it is found to hold as equality.

This algorithm, based on Theorem~\ref{t-maxnash}, for
enumerating all Nash equilibria of a rank-1 game has the
following noteworthy features.
First, it works for all games (degenerate or not), and its
characterization of maximal Nash subsets is simpler than for
general bimatrix games \citep{ARSvS}, and could even be
adapted to easily represent these Nash subsets in terms of
their inequalities rather than their vertices (which would
be of interest if they are high-dimensional).
Secondly, the algorithm in effect traverses $\N$ which is
generically a path.
Rather than by solving a succession of LPs, it can also be
implemented by a variant of the algorithm by
\citet{Lemke1965} with the additional linear constraints
$\lambda\ge x\T a$ or $\lambda\le x\T a$, depending on the
current sign of $\lambda-x\T a$.
Here, traversing this path gives \textit{all} Nash
equilibria, whereas for general bimatrix games Lemke's
algorithm (as in \citealp{vSET} or \citealp{GW03})
only finds \textit{one} Nash equilibrium.
\section{Two examples}
\label{s-example}

In this section, we illustrate the results of the previous
sections with an example of a rank-1 game.
After that we will give an example that shows that binary
search will in general not work for a game of rank~2 or
higher, even though Lemma~\ref{l-equiv} suggests the
possibility of finding a Nash equilibrium of such a game via
a recursive rank reduction.

Consider the following rank-1 game $(A,B)$,
\begin{equation}
\renewcommand{\arraystretch}{1}
\label{ex1}
A=\left[\,\begin{matrix}1~& 0 \\ 0~& 1 \\
\end{matrix}\,\right], 
\quad
B=\left[\begin{matrix}\hfill1 & -2 \\ -1 & \hfill0 \\
\end{matrix}\,\right], 
\quad 
A+B
=\left[\begin{matrix}\hfill2 & -2 \\ -1 & \hfill1 \\
\end{matrix}\,\right] 
=ab\T,
\end{equation}
where $a\T=(2,-1)$ and $b\T=(1,-1)$.
This game has the two pure equilibria
$((1,0),(1,0))$ and
$((0,1),(0,1))$,
and the mixed equilibrium $((\frac14,\frac34),(\frac12,\frac12))$.
By Theorem~\ref{t-equiv}(b), these are the equilibria
$(x,y)$ of the game $(A,-A+\1\lambda b\T)$ so that
$x\T a=\lambda$.
For $x=(1,0),(\frac14,\frac34), (0,1)$, this means
$\lambda=2,-\frac14,-1$.  

\begin{figure}[hbt]
\strut\hfill 
\includegraphics[width=.5\hsize]{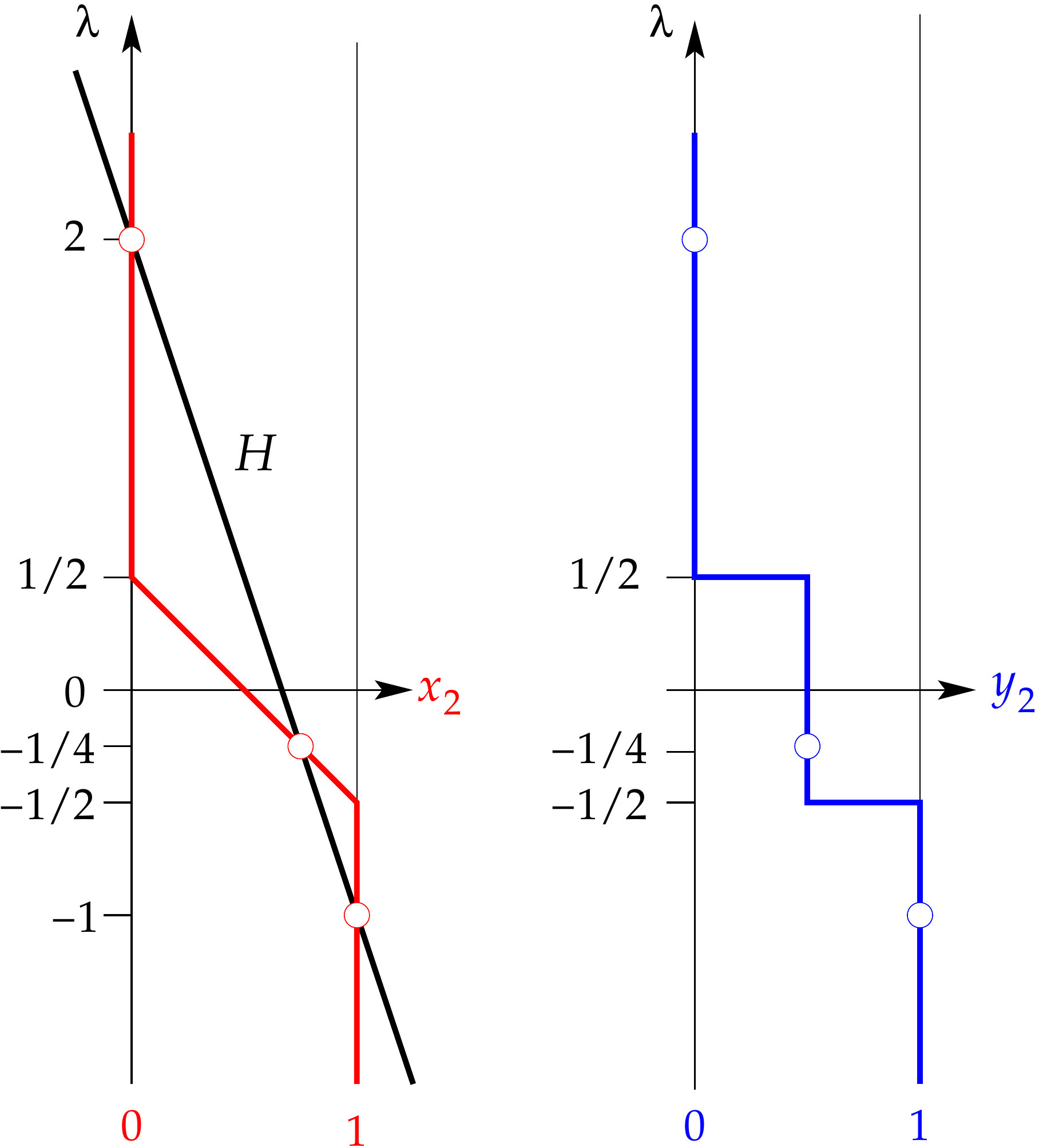}
\hfill 
\strut 
\caption{%
(Color online)
The path $\cal N$ in (\ref{NA}) for the game (\ref{N1}),
for $x=({1-x_2},x_2)\in X$ and $y=(1-y_2,y_2)\in
Y$, and the hyperplane $\H$ in~(\ref{H})}
\label{f1}
\end{figure}

Figure~\ref{f1} shows the set $\N$
in (\ref{NA}) where $(x,y)$ is an equilibrium of the
parameterized game $(A,-A+\1\lambda b\T)$, where 
\begin{equation}
\label{N1}
\renewcommand{\arraystretch}{1}
-A+\1\lambda b\T=
\left[\begin{matrix}-1~& \hfill0~\\ \hfill0~& -1~\\
\end{matrix}\right]
+
\left[\begin{matrix}
~\lambda & -\lambda \\
~\lambda & -\lambda \\
\end{matrix}\right].
\end{equation}
These equilibria are pure except when
$\lambda\in[-\frac12,\frac12]$,
when the unique mixed strategy $(1-x_2,x_2)$ of player~1
is given by equalizing the column payoffs,
$-(1-x_2)+\lambda=-x_2-\lambda$, that is,
$\lambda=\frac12-x_2$.
The white dots indicate
the intersection of $\N$ with the hyperplane~$H$
in~(\ref{H}), which is defined by the equation
$\lambda=x\T a=2(1-x_2)-x_2=2-3x_2$,
and no constraints on~$y$.

\begin{figure}[hbt]
\strut\hfill 
\includegraphics[width=.5\hsize]{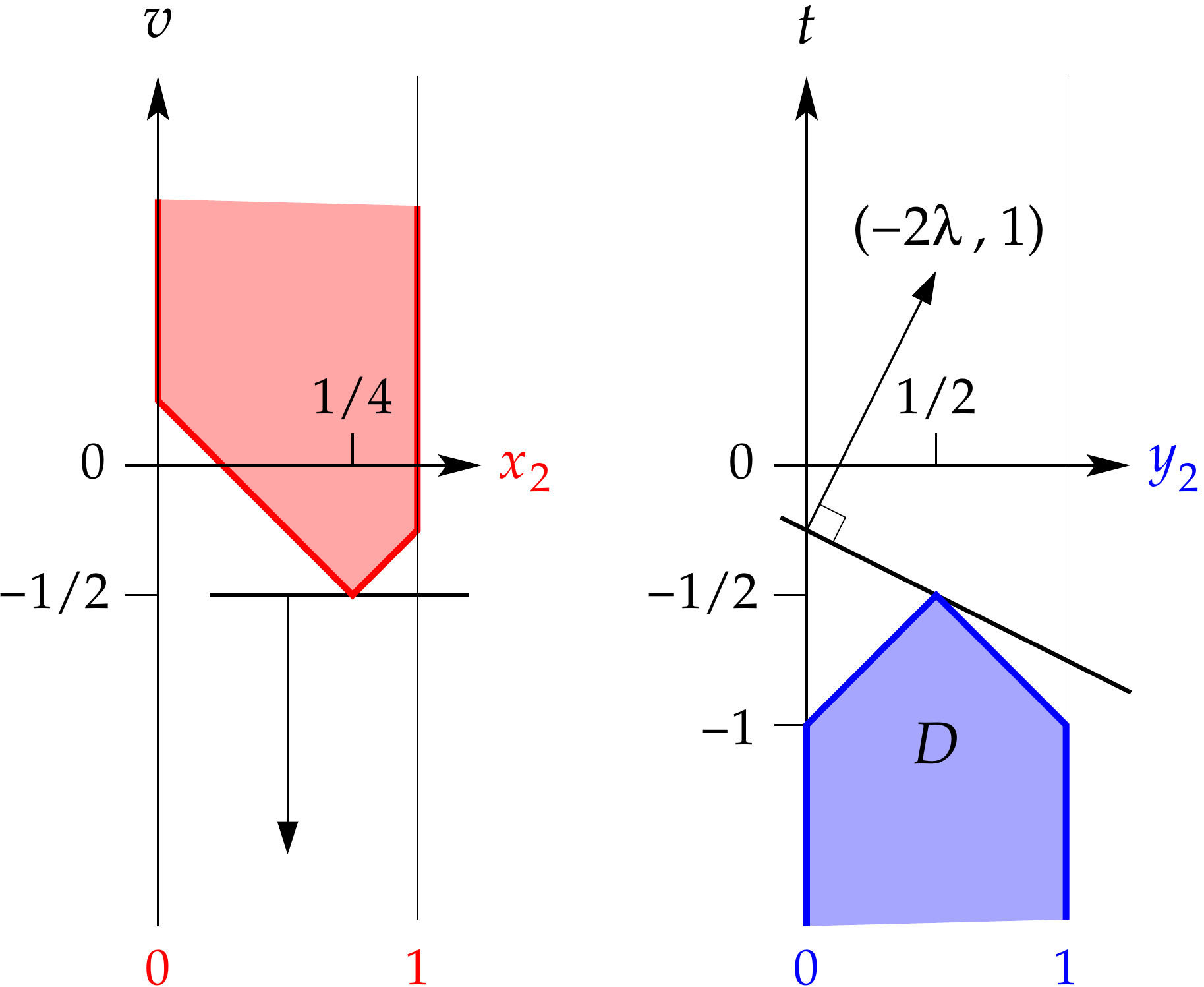}
\hfill 
\strut 
\caption{%
(Color online)
The LP $P_\lambda$ in
(\ref{Plamb}) and the polyhedron $D$ in (\ref{defD}) with the
objective function of the LP $D_\lambda$
in (\ref{Dlamb}) for $\lambda=-\frac14$, for the game (\ref{N1})
}
\label{f1lp}
\end{figure}

Figure~\ref{f1lp} shows
the domains of the LPs $P_\lambda$ in (\ref{Plamb}) 
and $D_\lambda$ in (\ref{Dlamb}) for $\lambda=-\frac14$.
Again we show $x$ in~$X$ as $(1-x_2,x_2)$ and
$y$ in~$Y$ as $(1-y_2,y_2)$.
The constraints of $P_\lambda$ are then
$1-x_2+v\ge\lambda$ and
$x_2+v\ge-\lambda$, which for $\lambda=-\frac14$ are
$v\ge-\frac54+x_2$ and
$v\ge\frac14-x_2$.
The constraints ${Ay+\1t}\le\0$ of $D_\lambda$ are
\begin{equation}
\label{C1}
1-y_2+t\le0\qquad\hbox{and}\qquad y_2+t\le0\,,
\end{equation}
and the objective function $\lambda b\T y+t$ is
$\lambda (1-y_2-y_2)+t$,
with gradient
$(\frac{\partial}{\partial y_2},\frac{\partial}{\partial
t})=(-2\lambda,1)=(\frac12,1)$ for $\lambda=-\frac14$.
For $\lambda>\frac12$, the optimum of $D_\lambda$ is
attained at the vertex $(y_1,y_2,t)=(1,0,-1)$ of $D$,
for $\frac12>\lambda>-\frac12$ at the vertex
$(\frac12,\frac12,-\frac12)$, and 
for $-\frac12>\lambda$ at the vertex $(0,1,-1)$.
For $\lambda_2=\frac12$ and $\lambda_1=-\frac12$, the optimal
face of $D_\lambda$ is an edge of $D$.
These are the two breakpoints $\lambda_1$ and
$\lambda_2$ in Theorem~\ref{t-endp}.

Figure~\ref{f1} also demonstrates the characterization of
the path $\N$ in Theorem~\ref{t-N}.
The left diagram shows (from left to right) the three pieces 
$X'_2$, $X'_1$, $X'_0$, each of which happen to
intersect~$\H$.
In the central diagram, the vertical parts of the path are
$Y'_2$, $Y'_1$, $Y'_0$, and the horizontal parts (for the
breakpoints) are $Y_2$ and $Y_1$.
This corresponds to the following, more elementary
game-theoretic explanation.
Except when $\lambda=-\frac12$ or $\lambda=\frac12$,
player~2's 
equilibrium strategy $y$ in
the game $(A,-A+\1\lambda b\T)$
is constant in $\lambda$, which holds because player~1's
payoff matrix $A$ does not change with~$\lambda$ and $y$ is
chosen so as to make player~1 indifferent between the pure
strategies in the support of his equilibrium strategy.
When $\lambda=-\frac12$ or $\lambda=\frac12$, the game is
degenerate, and player~2's equilibrium strategies form a
line segment, which allows the change of support of her
equilibrium strategy~$y$.

\begin{figure}[hbt]
\strut\hfill 
\includegraphics[width=90mm]{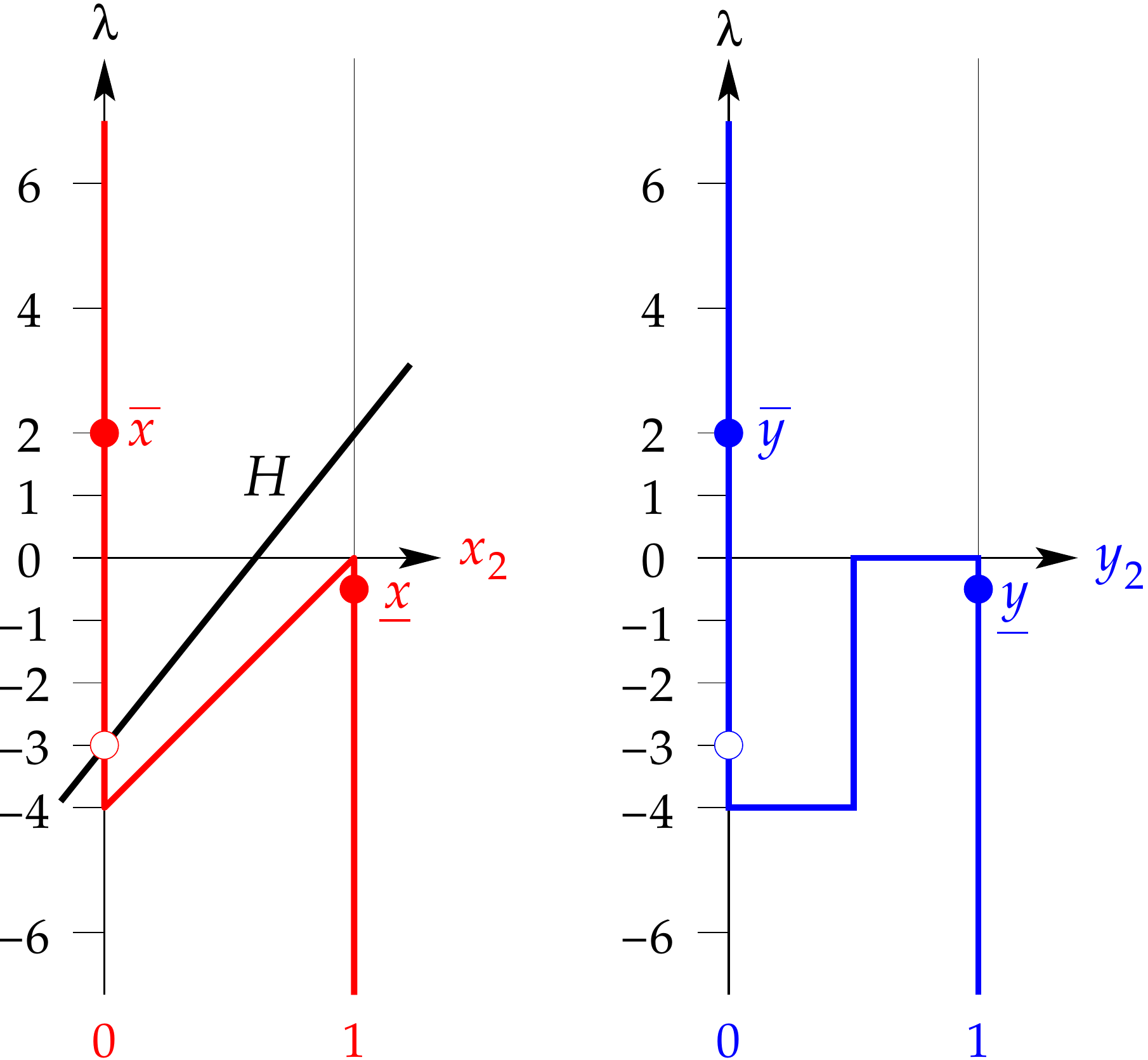}
\hfill 
\strut 
\caption{%
(Color online)
The path $\cal N$ of equilibria of the games in
(\ref{N2}) where the binary search method fails} 
\label{f2}
\end{figure}

Our second example shows that the binary search algorithm
no longer works for rank-$r$ games with $r>1$.  
Consider the following game $(A,B)$ of rank~2:
\begin{equation}
\label{ex3}
\renewcommand{\arraystretch}{1}
A=\left[\,\begin{matrix} 1 & -1 \\ 0 & \hfill 0\, \\
\end{matrix}\right], 
\quad
C=\left[\,\begin{matrix}4 & ~~0 \\ 0 & ~~0 \\ \end{matrix}\right],
\quad
B=C + a b\T=\left[\,\begin{matrix}1 & ~~0 \\ 2 & ~~0 \\
\end{matrix}\right], 
\end{equation} 
where $a\T=(-3,2)$ and $b\T=(1,0)$.
Here, $(A,B)$ is of rank 2 and $(A,C)$ is of rank 1.
The only equilibrium of $(A,B)$ is the pure equilibrium
$((1,0),(1,0))$.
The parameterized game $(A,C+\1\lambda b\T)$ has payoff
matrices 
\begin{equation}
\label{N2}
\renewcommand{\arraystretch}{1}
A=\left[\,\begin{matrix} 1 & -1 \\ 0 & \hfill 0\, \\
\end{matrix}\right], 
\quad
C+\1\lambda b\T=
\left[\,\begin{matrix} 
4+\lambda  &~~0 \\
\hfill \lambda & ~~0 \\
\end{matrix}\right].
\end{equation}
It has the following equilibria $(x,y)$ depending
on~$\lambda$, which define the set $\N$ in (\ref{N}), shown
in Figure~\ref{f2}:
The pure equilibrium $((1,0),(1,0))$ for ${\lambda\ge-4}$;
the pure equilibrium $((0,1),(0,1))$ for $\lambda\le0$;
the mixed equilibrium $((-\frac\lambda4,1+\frac\lambda4),
(\frac12,\frac12))$ for ${-4<\lambda<0}$,
and two further components
$((1,0),(1-y_2,y_2))$ with $y_2\in[0,\frac12]$ when $\lambda=-4$ and
$((0,1),(1-y_2,y_2))$ with $y_2\in[\frac12,1]$ when
$\lambda=0$ where the game in (\ref{N2}) is degenerate.
These are multiple disjoint equilibrium components for
$-4\le\lambda\le0$, which cannot happen for a parameterized
zero-sum game.
As a result, $\lambda$ may change non-monotonically along the
path $\N$, which in general causes a binary search to fail,
as we show next.

We describe a suitably adapted binary search method for this
example, where instead of solving parameterized LPs we find
equilibria of the parameterized game (\ref{N2}) of lower
rank.
The smallest and largest components of $a$ as in
line~3 of the \BINSEARCH{} algorithm are $\llam=-3$
and $\ulam=2$.
For $\lambda=\overline\lambda$, the only equilibrium of the game
in (\ref{N2}) is
$(\overline x,\overline y)=((1,0),(1,0))$, but for 
$\lambda=\underline\lambda$ there are multiple equilibria,
where we choose $(\underline x,\underline y)=((0,1),(0,1))$.
Then $\underline\lambda=-3<\underline x\T a=2$ and
$\overline x\T a=-3<\overline\lambda=2$, so
we next consider the midpoint $\lambda=(\llam+\ulam)/2=-1/2$
as in line~5 of \BINSEARCH, and compute
a new equilibrium of this parameterized game.
Suppose this is again $(x,y)=((0,1),(0,1))$, so that because
$\lambda<x\T a$ the assignment 
$(\underline\lambda,\underline x, \underline y)
\leftarrow(\lambda, x, y)$
takes place for the binary search to continue.
This is the situation shown in Figure~\ref{f2}.
At this point, the method will no longer succeed in finding
a suitable value of $\lambda$ because the search interval
$[\underline\lambda,\overline\lambda]=[-\frac12,2]$ no longer
contains the only possible value for $\lambda$, namely $-3$.
The problem is that in that interval, the set $\N$ consists
of two disconnected parts where $\lambda <x\T a$
and $\lambda >x\T a$ on opposite sides of the
hyperplane~$\H$, so that $\N$ no longer intersects with $\H$.
Hence, even though the values of $\lambda$ converge, the
corresponding equilibria $(x,y)$ on the two sides of $\H$
will not converge.

This example shows that because of the non-monotonicity of
$\lambda$ along the path $\N$, there is no equivalent
statement to Lemma~\ref{l-cont} that would guarantee that a
binary search will succeed.

\section{Rank-1 games with exponentially many equilibria}
\label{s-expo}

\citet[Open Problem~9]{KT2010} asked if the number of Nash
equilibria of a nondegenerate rank-1 game is polynomially
bounded.
This is not the case, because our next result shows that
this number may be exponential.

\begin{theorem}
\label{t-expo}
Let $p>2$ and let $(A,B)$ be the $n\times n$ bimatrix game
with entries of $A$ 
\begin{equation}
\label{aij}
a_{ij}=
\begin{cases}
2p^{i+j} & \hbox{if } j>i\\
p^{2i} & \hbox{if } j=i\\
0 & \hbox{if } j<i\\
\end{cases}
\end{equation}
for $1\le i,j\le n$, and $B=A\T$.
Then $A+B$ is of rank $1$, and $(A,B)$ is a nondegenerate
bimatrix game with $2^n-1$ many Nash equilibria.
\end{theorem}

\myproof
By (\ref{aij}), $A+B=a b\T$ with the $n$ components of
$a$ and $ b$ defined by $ a_i=p^i$ and $ b_j=2p^j$ for $1\le
i,j\le n$, so $A+B$ is of rank~1.

Let $y\in Y$ with support $S$.
Consider a row~$i$ and let $T=\{j\in S\mid j>i\}$.
Because $A$ is upper triangular, the expected payoff
against $y$ in row $i$ is
\begin{equation}
\label{Ayi}
(Ay)_i=a_{ii}y_i+\sum_{j\in T}a_{ij}y_j\,.
\end{equation}
Suppose $i\not\in S$.
If $T$ is empty, then $(Ay)_i=0<(Ay)_1$, otherwise 
let $t=\min T$ and note that for $j\in T$ we have
$a_{ij}=2p^{i+j}<p^{1+i+j}\le p^{t+j}\le a_{tj}$, so 
$(Ay)_i<(Ay)_t$.
Hence, no row~$i$ outside $S$ is a best response to $y$.
Similarly, because the game is symmetric, any column that is
a best response to
$x$ in $X$
belongs to the support of~$x$.
This shows that the game is nondegenerate.
Moreover, if $(x,y)$ is an equilibrium of $(A,B)$, then
$x$ and $y$ have equal supports.

For any nonempty subset $S$ of $\{1,\ldots,n\}$, we
construct a mixed strategy $y$ with support~$S$ so that
$(y,y)$ is an equilibrium of $(A,B)$.
This implies that the game has $2^n-1$ many equilibria,
one for each support set $S$.
The equilibrium condition holds if $(Ay)_i=u$ for $i\in S$
with equilibrium payoff~$u$, because then $(Ay)_i<u$ for
$i\not\in S$ as shown above.
We start with $s=\max S$, where $(Ay)_s=a_{ss}y_s=u$, by
fixing $u$ as some positive constant (e.g., $u=1$), which
determines $y_s$.
Once $y_i$ is known for all $i\in S$ (and $y_i=0$ for
$i\not\in S$), we scale $y$ and $u$ by multiplication with
$1/\1\T y$ so that $y$ becomes a mixed strategy.
Assume that $i\in S$ and $T=\{j\in S\mid j>i\}\neq\emptyset$
and assume that $y_k$ has been found for all $k$ in $T$ so
that $(Ay)_k=u$ for all $k$ in $T$, which is true for
$T=\{s\}$.
Then, as shown above,
$\sum_{j\in T}a_{ij}y_j<\sum_{j\in T}a_{tj}y_j=(Ay)_t=u$ for
$t=\min T$, so $y_i$ is determined by $(Ay)_i=u$ in
(\ref{Ayi}), and $y_i>0$.
By induction, this determines $y_i$ for all $i$ in $S$, and
after re-scaling gives the desired equilibrium strategy~$y$.  
\myendproof

By Theorem~\ref{t-equiv}, the equilibria $(x,y)$ of a
rank-1 game are the intersection of the path $\N$ in
(\ref{NA}) with the hyperplane $\H$ in (\ref{H}).
The exponential number of Nash equilibria of the game in
Theorem~\ref{t-expo} shows that $\N$ has exponentially many
line segments.
\citet{Murty1980} describes a parameterized LP with such an
exponentially long path of length $2^n$.
The payoffs for the game in Theorem~\ref{t-expo} have been
inspired by Murty's example, but are not systematically
constructed from it, which would be interesting.
See \cite{vS2012} for further discussions and related work
on the maximal number of Nash equilibria in bimatrix games,
such as \cite{vS1999}.

\section{A rank-preserving structure theorem}
\label{s-structure}

Nash equilibria of games are in general not unique, which
has led to a large literature on equilibrium
\textit{refinements} \citep{VD2} that impose additional
conditions on equilibria, such as \textit{stability} against
small changes in the game parameters, as proposed in the
seminal paper by \citet{KM} (KM).
They showed that stability has to apply to equilibrium
\textit{components}, that is, maximal sets of equilibria
that are topologically connected (which for bimatrix games
are unions of intersecting maximal Nash subsets, see
Section~\ref{s-enum}).
That is, an equilibrium component is stable if every
perturbed game has an equilibrium near that component
(although possibly in different positions depending on the
perturbation, which is why any single equilibrium may fail
to be stable).
KM proved the existence of stable equilibrium components
with the help of a \textit{structure theorem}
\cite[Theorem~1]{KM} which states that
the equilibrium correspondence $E$
over the set $\Gamma$ of strategic-form games
with a given number of players and numbers of strategies is
homeomorphic to $\Gamma$ itself.

In this section, we present in Theorem~\ref{t-hom} a similar
structure theorem with a new homeomorphism for bimatrix
games that \textit{preserves rank}.
In analogy to \citet[Appendix~B]{KM}, one consequence of
this new structure theorem 
is the existence of an equilibrium component in a game
$(A,B)$ that is stable with respect to small perturbations
that preserve the sum $A+B$ of the payoff matrices.
This is not interesting for zero-sum games which always have
only one component, but it is for games of higher rank and
applies, for example, to perturbations of the matrix~$A$ in
a rank-1 game given as $(A,-A+ab\T)$.
Furthermore, a number of equilibrium-finding algorithms can
be interpreted as following a path on the equilibrium
correspondence $E$ via the KM homeomorphism and suitable
projections \citep{wilson-computing,GW03}. 
As a topic for further research, it may be interesting to
study our new homeomorphism in this context, or, similar to
\cite{jansen2001computation}, the computation of equilibrium
components that are stable 
with respect to small perturbations
that preserve the sum $A+B$ of the payoff matrices.

We first recall the KM homeomorphism from \cite{KM}.
Let $\Gamma$ be the set of $m\times n$ bimatrix games
$(A,B)$ and $E\subseteq\Gamma\times X\times Y$ be its
equilibrium correspondence,
\begin{equation}
\label{corresp}
E=\{(A,B,x,y)\mid (A,B)\in\Gamma,~(x,y)\hbox{ is a NE of
}(A,B)\}.
\end{equation}
To distinguish the dimensions of the all-zero and all-one
vectors we write them as $\zerom,\onem\in\reals^m$ and
$\0,\onen\in\reals^n$.
Let $a$ and $b$ be the vectors of row and column averages of
$A$ and $B$,
\begin{equation}
\label{ab}
\textstyle
a=A\onen\frac1n,
\qquad
b=B\T\onem\frac1m\,.
\end{equation}
Then $A$ and $B$ correspond uniquely to pairs $(\tilde A,a)$
and $(\tilde B,b)$ with
\begin{equation}
\label{AaBb}
A=\tilde A+a\onen\T,
\qquad
B=\tilde B+\onem b\T,
\qquad
\tilde A\onen=\zerom,
\qquad
\onem\T\tilde B=\0\T, 
\end{equation}
with $a$ and $b$ as in~(\ref{ab}).
That is, $(A,B)$ is parameterized by a ``base game''
$(\tilde A,\tilde B)$ where each row of player~1 and each
column of player~2 gets payoff zero when the other player
randomizes uniformly (as in 
$\tilde A\onen\frac1n=\zerom$, where the factor $\frac1n$
does not matter), and a pair of vectors
$a$ in $\reals^m$ and $b\T$ with 
$b$ in $\reals^n$ that are added to the rows of $\tilde A$
and columns of~$\tilde B$, respectively, to obtain the
correct payoffs.

The KM homeomorphism $\phi:\Gamma\to E$ only changes $a$
and $b$.
It is most easily described by its inverse
$\phi^{-1}:E\to\Gamma$ defined by
$\phi^{-1}(A,B,x,y)=(C,D)$,
\begin{equation}
\label{CD}
C = \tilde A+(Ay+x)\onen\T,
\qquad
D = \tilde B+\onem(x\T B+y\T).
\end{equation}
That is, $(C,D)$ has the same ``base game''
$(\tilde A,\tilde B)$ as $(A,B)$ but different parameters
$(Ay+x)\in\reals^m$ and $(B\T x+y)\in\reals^n$.
The fact that $(x,y)$ is an equilibrium of $(A,B)$
implies that $\phi^{-1}$ is injective (and therefore $\phi$
well-defined), by the following intuition. 
Because $x$ is a best response to $y$, each row of the
vector $Ay$ of expected payoffs in the support of $x$ has
maximal and equal value $u$ among all components of $Ay$,
by (\ref{xbr}).
This condition allows us to re-construct $x$ from the
sum $c=Ay+x$, which is used in the definition of $C$ in
(\ref{CD}) and which can be obtained from~$C$.
Suppose the components $c_i$ of $c$ are heights of $m$
``poles in the water'' of which a certain amount $x_i$ is
``above the waterline'' depending on the ``water
level''~$w$, where
\begin{equation}
\label{water-x}
x_i 
=\max(c_i-w,0)\,,
\end{equation}
so $x_i\ge0$ and if $c_i<w$ then $x_i=0$.
For any $c\in\reals^m$, there is a unique choice of
$w\in\reals$ in~(\ref{water-x}) so that $\sum_{i=1}^m x_i=1$
and therefore $x\in X$.
By this construction of $w$ and $x$, all components $p_i$ of
the vector $p=c-x$ fulfill (a) $w=\max_k p_k$,
and (b) $x_i>0$ implies $p_i=w$, as when $p=Ay$ and $x$ is a
best response to~$y$.
In a similar way, $y$ is a best response to $x$ and the sum
$x\T B+y\T$ used to define $D$ in (\ref{CD}) is special
because it allows us first to obtain a vector $d\in\reals^n$
from $D$, and second to obtain the original $y\in Y$ and
$q\in\reals^n$ so that $d=q+y$ and $q\T=x\T B$.
The following lemma states this construction, which we apply
afterwards to define the KM homeomorphism, and will later
use again for our new homeomorphism.

\begin{Lemma}
\label{l-recover}
Given $c\in\reals^m$ and $d\in\reals^n$, there are unique
$x\in X$, $y\in Y$, $p\in\reals^m$ and $q\in\reals^n$ so
that
\begin{equation}
\label{br}
\arraycolsep.2em
\begin{array}{rclrclll}
c&=&p+x\,,&
&d&=&q+y\,,
\\
x_i&=&0&
\strut\hbox{or}\qquad\strut
&
\displaystyle
p_i&=&u=
\max_{1\le k\le m}p_k
&\qquad(1\le i\le m),\\
y_j&=&0&
\strut\hbox{or}\qquad\strut
&
\displaystyle
q_j&=&v
=\max_{1\le l\le n}q_l
&\qquad(1\le j\le n).\\
\end{array}
\end{equation}
\end{Lemma} 

\myproof
For $t\in\reals$, let $t^+=\max(t,0)$, and
\begin{equation}
\label{waterlevel}
\begin{array}{rcl}
u&=&\min\{\,w\in\reals \mid \sum_{i=1}^m (c_i-w)^+\le 1\,\},
\\
v&=&\min\{\,w\in\reals \mid \sum_{j=1}^n (d_j-w)^+\le 1\,\},
\end{array}
\end{equation}
where $u$ (and similarly $v$) is the unique
lowest ``water level'' $w$ so that the ``heights'' of the
components $c_i$ of $c$ that are ``above the waterline'' sum
up to (at most) one.
Then
\begin{equation}
\label{getxy}
x_i=(c_i-u)^+
\qquad(1\le i\le m),
\qquad
y_j=(d_j-v)^+
\qquad(1\le j\le n),
\end{equation}
and $p=c-x$ and $q=d-y$ fulfill~(\ref{br}), and $x,y,p,q$
are uniquely determined by the conditions $x\in X$, $y\in
Y$, and~(\ref{br}).
\myendproof

The KM homeomorphism $\phi:(C,D)\mapsto(A,B,x,y)$ is then
defined as follows. 
{\myitem{(a)}
Let $c=C\onen\frac1n$,
$d=D\T\onem\frac1m$, 
$\tilde A=C-c\onen\T$ and $\tilde B=D-\onem d\T$.

\myitem{(b)}
Apply Lemma~\ref{l-recover} to get $x,y,p,q$ so that
(\ref{br}) holds.

\myitem{(c)}
Let $a=
c-x-\tilde Ay$ and $b=d-y-\tilde B\T x$, and define $A$ and $B$
by (\ref{AaBb}).
}

\noindent
Then $\phi$ is continuous because it is defined by
continuous linear mappings and (\ref{waterlevel}) and
(\ref{getxy}) for~(b).
We show that $(A,B,x,y)\in E$.
We have
$Ay=(\tilde A+a\onen\T)y=\tilde Ay+a=\tilde Ay+c-x-\tilde A
y=p$, and similarly $x\T B=x\T\tilde B+b\T=d\T-y\T=q\T$.
Then the conditions (\ref{br}) are equivalent to the
best-response conditions (\ref{xbr}) and (\ref{ybr}), that
is, $(x,y)$ is indeed an equilibrium of $(A,B)$.
Moreover, $c=p+x=Ay+x$ and $d=B\T x+y$, which shows that
the (continuous) function $(A,B,x,y)\mapsto (C,D)$ in
(\ref{CD}) is indeed the inverse of $\phi$ (so $\phi$ is
injective), and also that $\phi$ is surjective, because we
can start in (\ref{CD}) from any $(A,B,x,y)\in E$.

The KM homeomorphism does not operate within a subset of
games of fixed rank (for example, the zero-sum games).
Our new homeomorphism $\psi:\Gamma\to E$ has this property.
Consider a fixed matrix $M\in\reals^{m\times n}$, the set
$\Gamma_M$ bimatrix games $(A,B)$ with $A+B=M$, and $E_M$ as
the equilibrium correspondence $E$ in (\ref{corresp})
restricted to these games, 
\begin{equation}
\label{GM}
\Gamma_M=\{ (A,B)\in\Gamma\mid A+B =M\},
\qquad
E_M=\{ (A,B,x,y)\in E\mid(A,B)\in\Gamma_M\}.
\end{equation}
The following theorem states we can restrict
$\psi$ to a homeomorphism $\Gamma_M\to E_M$ for any $M$
(for example, the all-zero matrix $M$).
Also, $\psi$ is continuous in $M$ and therefore a
homeomorphism $\Gamma\to E$ like the KM homeomorphism.

\begin{theorem}
\label{t-hom}
Let $M\in\reals^{m\times n}$.
There is a homeomorphism $\psi:\Gamma_M\to E_M$,
$(C,D)\mapsto(A,B,x,y)$, that is, $A+B=M$ for 
all $(C,D)\in\Gamma_M$.
\end{theorem}

\myproof
We will use a new parameterization of any matrix $A$ in
$\reals^{m\times n}$, which
corresponds uniquely to a quadruple
$(\hat A,\gamma,a,b)$ with $\hat A\in\reals^{m\times n}$,
$\gamma\in\reals$, $a\in\reals^m$, and $b\in\reals^n$
according to 
\begin{equation}
\label{Agab}
A=\hat A+\onem\gamma\onen\T+a\onen\T+\onem b\T
\end{equation}
so that
\begin{equation}
\label{abcons}
\onem\T\hat A=\zeron\T,
\qquad
\hat A\onen=\zerom,
\qquad
\onem\T a=0,
\qquad
b\T\onen=0\,. 
\end{equation}
It is easy to see that $\hat A$, $\gamma$, $a$, and $b$ are
uniquely given by $A$, (\ref{Agab}), and
\begin{equation}
\label{howab}
\textstyle
\gamma=\frac1m\onem\T A\onen\frac1n,
\qquad
a=A\onen\frac1n-\onem\gamma,
\qquad
b\T=\frac1m\onem\T A-\gamma\onen\T\,. 
\end{equation}
The homeomorphism $\psi:\Gamma_M\to E_M$,
$(C,D)\mapsto (A,B,x,y)$ uses this parameterization of $C$
and only changes the vectors $a$ and $b$, and maintains
the sum $M$ of the payoff matrices, that is, $A+B=C+D=M$.
Like for the KM homeomorphism, we first describe its inverse
$\psi^{-1}$, which maps $(A,B,x,y)$ in $E_M$ to $(C,D)$ in
$\Gamma_M$.
Let $A+B=M$ and $(x,y)$ be an equilibrium of $(A,B)$.
Let $A$ be represented as in (\ref{Agab}) so that
(\ref{abcons}) holds, and let
\begin{equation}
\label{Cgcd}
C=\hat A+\onem\gamma\onen\T+c\onen\T+\onem d\T
\end{equation}
with $c$ and $d$ given by
\begin{equation}
\label{cd}
c=\rho(Ay+x),
\qquad
d=\sigma(B\T x+y)
\end{equation}
where $\rho:\reals^m\to\reals^m$ and
$\sigma:\reals^n\to\reals^n$ are the linear projections on the
hyperplane through the origin with normal vector $\onem$
respectively $\onen$,
\begin{equation}
\label{rhosigma}
\textstyle
\rho(x)=x-\onem(\frac1m\onem\T x),
\qquad
\sigma(y)=y-\onen(\frac1n\onen\T y)
\end{equation}
which achieves $\onem\T\rho(x)=0$ and $\onen\T\sigma(y)=0$ for any
$x\in\reals^m$ and $y\in\reals^n$, as required for a
parameterization of the payoff matrix $C$ like it is done for
$A$ in (\ref{abcons}).
With $C$ thus encoded, we let $D=M-C$.

The homeomorphism $\psi:(C,D)\mapsto(A,B,x,y)$ itself is
obtained as follows.
Let $(C,D)\in\Gamma_M$.
Similar to (\ref{Agab}) we represent $C$ by (\ref{Cgcd}) 
where as in (\ref{howab}) 
\begin{equation}
\label{howc}
\textstyle
\gamma=\frac1m\onem\T C\onen\frac1n\,,
\qquad
c=C\onen\frac1n-\onem\gamma\,,
\qquad
d\T=\frac1m\onem\T C-\gamma\onen\T, 
\end{equation} 
which implies
\begin{equation}
\label{cdcons}
\onem\T\hat A=\zeron\T,
\qquad
\hat A\onen=\zerom\,,
\qquad
\onem\T c=0\,,
\qquad
d\T\onen=0\,.
\end{equation}
Given $c$ and $d$, we determine $x\in X$, $y\in Y$,
$p\in\reals^m$ and $q\in\reals^n$ by
Lemma~\ref{l-recover} so that (\ref{br}) holds.
Then, let 
\begin{equation}
\label{abnew}
a=c-\rho(\hat Ay+x),
\qquad
b=\sigma((M-\hat A)\T x+y)-d
\end{equation}
so that $a$ and $b$ fulfill (\ref{abcons}),
define $A$ by (\ref{Agab}), and let $B=M-A$.
Like $\phi$ before, $\psi$ is defined by linear maps and the
continuous operations in (\ref{waterlevel}) and
(\ref{getxy}) and is therefore continuous.

We show that $\psi(C,D)=(A,B,x,y)\in E_M$.
Because $A+B=M$, we only need to show the equilibrium
property.
Using
(\ref{Agab}),
$\onen\T y=1$,
(\ref{abnew}),
$c=p+x$,
and the definition of $\rho$ in
(\ref{rhosigma}),
\begin{equation}
\label{Ay}
\textstyle
\arraycolsep.2em
\begin{array}{rcl}
Ay
&=&
\hat Ay +\onem\gamma\onen\T y +a\onen\T y+\onem b\T y
\\
&=&
\hat Ay +\onem\gamma+a +\onem b\T y
\\
&=&
\hat Ay +\onem\gamma+ c-\rho(\hat Ay+x) +\onem b\T y
\\
&=&
\hat Ay +\onem\gamma+ p+x
-(\hat Ay+x)
+\onem(\frac1m\onem\T (\hat Ay+x))
+\onem b\T y
\\
&=&
p+\onem(\gamma+\frac1m\onem\T (\hat Ay+x)+b\T y)
\\
&=&
p+\onem\alpha
\\
\end{array}
\end{equation}
for some $\alpha\in\reals$ 
which means that $(Ay)_i=p_i+\alpha$ for $1\le i\le m$
and therefore by (\ref{br}) the best-response condition
(\ref{xbr}) holds (which is unaffected by a constant shift),
that is, $x$ is a best response to~$y$.
Similarly, using $\onem\T x=1$,
(\ref{abnew}),
the definition of $\sigma$ in (\ref{rhosigma}),
and
$d=q+y$,
\begin{equation}
\label{Bx}
\textstyle
\arraycolsep.2em
\begin{array}{rcl}
B\T x
&=&
(M-A)\T x
\\
&=&
(M-\hat A-\onem\gamma\onen\T -a\onen\T-\onem b\T)\T x
\\
&=&
(M-\hat A)\T x
-\onen\gamma\onem\T x -\onen a\T x-b\onem\T x
\\
&=&
(M-\hat A)\T x
-\onen\gamma -\onen a\T x-b
\\
&=&
(M-\hat A)\T x
-\onen\gamma -\onen a\T x-
\sigma((M-\hat A)\T x+y)+d
\\
&=&
-\onen\gamma -\onen a\T x- y
+\onen\frac1n\onen\T ((M-\hat A)\T x+y)+q+y
\\ 
&=&
\onen\beta +q
\\
\end{array}
\end{equation}
for some $\beta\in\reals$ 
which means that $(B\T x)_j=q_j+\beta$ for $1\le j\le n$
and therefore by (\ref{br}) the best-response condition
(\ref{ybr}) holds, that is, $y$ is a best response to~$x$.
Hence, $(x,y)$ is indeed an equilibrium of $(A,B)$.

To show that $\psi$ has the inverse described in
(\ref{Cgcd}) and (\ref{cd}), note that $\rho$ and $\sigma$
in (\ref{rhosigma}) are linear and $\rho(\onem)=\zerom$ and
$\sigma(\onen)=\zeron$.
Therefore, for $\psi(C,D)=(A,B,x,y)$ with $C$ as in
(\ref{Cgcd}), we have by (\ref{Ay}) and (\ref{Bx}) 
and because $\onem\T c=\zerom$ and $\onen\T d=\zeron$,
\begin{equation}
\label{psi}
\textstyle
\arraycolsep.2em
\begin{array}{rclclclcl}
\rho(Ay+x)
&=&
\rho(p+\onem\alpha+x)
&=&
\rho(p+x)
&=&
\rho(c)
&=&
c\,,
\\
\sigma(B\T x+y)
&=&
\sigma(\onen\beta+q+y)
&=&
\sigma(q+y)
&=&
\sigma(d)
&=&
d\,,
\\
\end{array}
\end{equation}
that is, $\psi$ has indeed the (continuous) inverse
described in (\ref{cd}) and $\psi$ is both injective and
surjective.
This shows that $\psi$ is indeed a homeomorphsim from
$\Gamma_M$ to~$E_M$.
\myendproof

\section{Conclusions}
\label{s-conclusions}

We conclude with some open questions.
Our analysis shows that rank-1 games are computationally
easy to analyze:
One Nash equilibrium can be found in polynomial time, and
enumerating all equilibria can be performed by following a
piecewise linear path, similar to finding a single Nash
equilibrium of a bimatrix game (which is in general a
PPAD-hard problem).

As described in Section~\ref{s-enum}, the path of solutions
to the parameterized LP consists in general of polyhedral
segments whose intersections with the hyperplane $\H$
define the sets of Nash equilibria of the rank-1 game.
This set-up suggests the application of \textit{smoothed
analysis} as pioneered by \citet{ST2004} for the ``shadow
vertex algorithm'' for parameterized LPs.
This analysis has been subsequently improved and simplified;
for recent developments see \cite{DH2018}. 
In smoothed analysis, the LP data are perturbed by some
moderate Gaussian noise which cancels ``pathological'' cases
that lead to exponential worst-case examples, like the game
constructed in Section~\ref{s-expo}.
Applied to our parameterized LP, it would imply that in
expectation there is a polynomial number of segments in
Theorem~\ref{t-N}.
If this holds, the number of Nash equilibria is similarly
polynomially bounded by Theorem~\ref{t-maxnash} (the Nash
subsets are all single equilibria because the perturbed game
is generic and therefore nondegenerate with probability
one).
However, the standard framework of smoothed analysis (as in
e.g.\ \citealp{DH2018}) assumes that the LP constraints are of
the form $Ax\le\1$, which is not the case for the LP
(\ref{lp1}) that we consider, so combining this with our
approach requires a careful study that we leave for future
work.
For a general bimatrix game, finding one equilibrium is
PPAD-hard even under smoothed analysis \citep{CDT2009}.
However, it is not known if a perturbed game may have
exponentially long Lemke--Howson paths; the long paths in
\cite{SvS2006} do not persist due to exponential size
differences in the payoffs.

In Section~\ref{s-expo} we described rank-1 games with
exponentially many equilibria (also with exponential size
differences in the payoffs).
This raises the following question:
Can all equilibria of a rank-1 game be computed in running
time that is polynomial in the size of the input
\textit{and output}?
Such an algorithm is called ``output efficient''. 
For example, the algorithm by \citet{AdlerM1992} that
computes all segments of a parameterized LP is output
efficient.
We have extended this algorithm in Section~\ref{s-enum}.
For general bimatrix games, an output efficient algorithm
that finds all Nash equilibria would imply P${}={}$NP
because it is NP-hard to decide if a game has more than one
Nash equilibrium \citep{GZ89}.
Our binary search algorithm gives no information about the
existence of a second equilibrium, so it is conceivable that 
finding a second Nash equilibrium of a rank-1 game is also
NP-hard.
The existence of an output efficient algorithm to find all
Nash equilibria of a rank-1 game is an open question.

General bimatrix games are computationally difficult, but
rank-1 games are computationally easy. 
One should therefore investigate \textit{economic
applications} of large rank-1 games, also as approximate
economic models that can serve as fast-solvable benchmarks.
As a possible starting point, we describe here a simple
``trade game'', which suggests that rank-1 games are
much more versatile and economically interesting than
zero-sum games.
Let player~1 be a seller of a product who can choose
possible \textit{quality levels} $a_i$ for $i=1,\ldots,m$,
and let player~2 be a buyer who can decide on possible
\textit{quantity levels} $b_j$ for $j=1,\ldots,n$ that she
buys from the seller.
A \textit{price} $p_{ij}$ that is paid from buyer to seller
can be chosen arbitrarily for each $i$ and~$j$.
Suppose there are further parameters $\alpha$, $\beta$,
$\gamma_j$, and $\delta_i$ so that the payoffs to the players are 
\begin{equation}
\label{P1}
\arraycolsep0pt
\begin{array}{rrl}
\hbox{payoff to player 1} :&
p_{ij}&{}-\alpha a_i b_j+\gamma_j
\\
\hbox{payoff to player 2} :&
~~{-}p_{ij}&{}+\beta a_i b_j+\delta_i 
\,.
\end{array}
\end{equation}
We further assume that $\beta>\alpha>0$, which
reflects that high quality is costly to produce for player~1
and beneficial for player~2, with $\beta-\alpha$
representing the benefits from trade.
The additional parameter $\gamma_j$ (increasing with $b_j$)
is an additional benefit to player~1 for higher amounts of
sold quantities, and similarly $\delta_i$ to player~2 for
higher quality.
Neither $\gamma_j$ nor $\delta_i$ affect the players' best
responses and can therefore assumed to be zero.
This gives a strategically equivalent game whose sums of
payoffs are $(\beta-\alpha)a_ib_j$ and therefore of rank
one.
Because rank-1 games can be analyzed very fast, this
``trade game'' can be studied for large values of $m$ and
$n$, and in particular for its possibly many price levels.
The concrete economic interpretation of such games and their
equilibria remains to be investigated. 
{\citet{BL2006} consider a ``multiplication game'' which is a
matching game between $n$ workers and $n$ firms where the 
suitability of a worker for a firm is described by a matrix
of rank one.
However, it is a game with $2n$ players, not two players.}

\section*{Acknowledgments}
This material is based upon work supported by the National
Science Foundation under Grant No.\ CRII 1755619
(Jugal Garg) and CAREER award CCF 1750436 (Ruta Mehta).

We thank two anonymous referees of \textit{Operations
Research} for their detailed comments
which helped improve the manuscript.
Heinrich Nax suggested the ``trade game'' (\ref{P1}) in
Section~\ref{s-conclusions}.

\addcontentsline{toc}{section}{References} 
\bibliographystyle{Eig}
\bibliography{bib-rank1} 

\end{document}